\documentclass[twocolumn,showpacs,preprintnumbers,amsmath,amssymb]{revtex4}

\usepackage{graphicx}
\usepackage{dcolumn}
\usepackage{bm}

\newcommand{\bq}{\begin{equation}}
\newcommand{\eq}{\end{equation}}
\newcommand{\bqa}{\begin{eqnarray}}
\newcommand{\eqa}{\end{eqnarray}}
\newcommand{\nn}{\nonumber \\}

\def\be     {\begin{equation}}
\def\ee     {\end{equation}}
\def\bea        {\begin{eqnarray}}
\def\eea        {\end{eqnarray}}
\def\bnn    {\begin{eqnarray*}}
\def\enn    {\end{eqnarray*}}

\begin{document}

\title{Fermionization for charge degrees of freedom and
bosonization of spin degrees of freedom in the SU(2) slave-boson
theory}
\author{Ki-Seok Kim}
\affiliation{Institute de Physique Th\'eorique, CEA, IPhT, CNRS,
URA 2306, F-91191 Gif-sur-Yvette, France}
\date{\today}

\begin{abstract}
Fermionizing the charge sector and bosonizing the spin part in the
SU(2) slave-boson theory, we derive an effective field theory for
dynamics of doped holes in the antiferromagnetically correlated
spin background, where spin fluctuations are described by an SO(5)
Wess-Zumino-Witten (WZW) theory while dynamics of doped holes is
characterized by QED$_{3}$ with a chemical potential term. An
important feature of our effective field theory is the coupling
term between valance bond fluctuations and doped holes.
Considering that valance bond fluctuations are deeply related with
monopole excitations of staggered U(1) gauge fields in the bosonic
field theory for spin fluctuations, we demonstrate that hole
dynamics helps deconfinement of bosonic spinons near the quantum
critical point of the SO(5) WZW theory. We solve this effective
field theory in the Eliashberg framework, and find non-Fermi
liquid physics in thermodynamics and transport, where $z = 3$
criticality with dynamical exponent $z$ plays an important role
for hole dynamics. We discuss validity of our field theory,
applying it to a doped spin chain and comparing it with the
slave-fermion framework. Furthermore, we discuss instability of
the anomalous metallic phase against superconductivity and density
waves of doped holes, resulting from competition between gauge and
valance bond fluctuations.
\end{abstract}


\maketitle

\section{Introduction}

Doping to an antiferromagnetic Mott insulator has been one of the
central interests in modern condensed matter physics, associated
with high T$_{c}$ superconductivity. Since the normal state is
extremely anomalous, particulary shown from the absence of
quasiparticle excitations near $(\pm \pi, 0)$ and $(0, \pm \pi)$
momentum points\cite{ARPES} and temperature quasi-linear behavior
in transport experiments\cite{Transport} although well defined
electron excitations seem to exist near $(\pm \frac{\pi}{2}, \pm
\frac{\pi}{2})$\cite{ARPES,Transport}, emergence of rather
conventional BCS-type superconductivity from such an abnormal
normal state has been a long standing puzzle and still does. Such
an anomalous behavior would be surely due to strong correlations
between electrons, associated with Mott physics.


Slave-boson approach has been one of the canonical frameworks for
study of strongly correlated electrons. In particular, a doped
Mott insulator problem was formulated in the slave-boson
context,\cite{Lee_Nagaosa_Wen} where strong repulsive interactions
cause so called the single occupancy constraint naturally imposed
in the slave-boson representation, and link variables arise as
collective "order parameter" excitations formulated as gauge
fields. U(1) slave-boson gauge theory has been enjoyed both
intensively and extensively for the doped Mott insulator problem,
but such a formulation turns out to have fundamental difficulty
for d-wave superconductivity emerging from a doped Mott
insulator.\cite{U1_Superfluid}

Wen and Lee have developed an SU(2) formulation, which extends the
U(1) slave-boson theory to include fluctuations between nearly
degenerate U(1) mean-field states, well applicable in underdoped
regions.\cite{Lee_Nagaosa_Wen} Based on their SU(2) construction,
they could obtain the doping independent decreasing ratio of
superfluid weight with a confinement ansatz. In addition, they
predicted a special structure of a vortex, which has a staggered
flux core. Furthermore, they argued similarity between their
staggered flux phase of the SU(2) slave-boson theory and the
Gutzwiller projected BCS wave function based on an explicit
numerical evaluation.

Although the SU(2) slave-boson framework has explained many kinds
of aspects for high T$_{c}$ cuprates such as phase diagram,
thermodynamics, transport, and etc.,\cite{Lee_Nagaosa_Wen}
antiferromagnetic spin fluctuations are difficult to take into
account in this context. High T$_{c}$ superconductivity is
sometimes argued to emerge from the spin liquid phase described by
the spin sector of the slave-boson theory in the slave-boson
community, but its connection to antiferromagnetism is an
important issue since the original problem is doping to an
antiferromagnetic Mott insulator instead of doping to the spin
liquid one. This motivates us to propose how to introduce
antiferromagnetic correlations in the SU(2) slave-boson framework.

Study of quantum antiferromagnetism associated with high T$_{c}$
cuprates has been performed in the context of spin liquid. In the
bosonic representation of spin the half filled quantum
antiferromagnet on the square lattice is described by the O(3)
nonlinear $\sigma$ model, allowing a quantum phase transition from
an antiferromagnet to a quantum disordered paramagnet. Bernevig et
al. have claimed that although the appropriate off-critical
elementary degrees of freedom are given by either spin $1$
excitons (gapped paramagnons) in the quantum disordered paramagnet
and spin $1$ antiferromagnons in the antiferromagnet, at the
quantum critical point such excitations should break up into more
elementary spin $1/2$ excitations usually called
spinons.\cite{Laughlin} This was challenged by Senthil et
al.\cite{Senthil_DQCP} They argued that since the phase transition
in Ref. \cite{Laughlin} is supposed to fall into
Landau-Ginzburg-Wilson paradigm, the spectrum at the quantum
critical point should be fully understandable only in terms of
spin $1$ bosonic degrees of freedom.\cite{Kim_DQCP} Senthil et al.
proposed, as a possible candidate for a deconfined quantum
critical point, a direct quantum phase transition between a Neel
antiferromagnet and a valance bond solid state, where one gets
spinon condensation in the Neel state while instanton excitations
(tunnelling events between energetically degenerate but
topologically inequivalent gauge vacua in the CP$^1$
representation of the O(3) nonlinear $\sigma$ model) should
possibly arise in the paramagnetic phase, whose condensation does
not allow spinon deconfinement. Senthil et al. demonstrated that
such proliferation of instantons is not the case at the quantum
critical point, where a topological $\theta$ term usually referred
as a Berry phase term makes instantons irrelevant and accordingly,
makes it possible to achieve spinon deconfinement.

This proposal motivated direct numerical simulations of various
microscopic models to find such an exotic quantum critical point
beyond the Landau-Ginzburg-Wilson paradigm. Actually, Sandvik has
claimed that such a critical point exists indeed in a modified
Heisenberg model with four-spin interactions.\cite{Sandvik}
Furthermore, he pointed out an important thing, that is, only one
length scale seems to exist and accordingly, critical exponents
for both staggered spin correlations and valance bond fluctuations
are the same as each other, apparently in contrast with the
original proposal for deconfined quantum criticality of the
bosonic field theory.

Resorting to the fermion representation of spin, QED$_{3}$ or
QCD$_{3}$ was obtained as an effective field theory at half
filling, where its conformal invariant fixed point is allowed in
the large $N$ limit, identified with algebraic spin
liquid.\cite{ASL_Mother,Wen_Symmetry} An important point is that
this effective field theory has an enlarged symmetry compared with
its microscopic Hamiltonian. QCD$_{3}$ resulting from the $\pi$
flux ansatz has Sp(4)$\approx$SO(5) while QED$_{3}$ arising from
the staggered flux gauge exhibits SU(4), where both are basically
associated with spin and nodal
structures.\cite{ASL_Mother,Wen_Symmetry} This enlarged symmetry
has important physical implication that symmetry equivalent
operators have their same scaling dimension, which should be taken
into account for an effective field theory of such composite field
variables.

Actually, Tanaka and Hu have derived an effective field theory for
competition between antiferromagnetism and valance bond
solid.\cite{Tanaka_SO5} Starting from the $\pi$ flux phase, they
obtain QCD$_{3}$ as its effective field theory although they do
not take such gauge fluctuations into account explicitly. Their
crucial observation is that the antiferromagnetic order parameter
is symmetry equivalent to the valance bond one via chiral rotation
in SO(5). Introducing a chiral rotated mass term with an SO(5)
superspin vector, they could derive an SO(5) Wess-Zumino-Witten
(WZW) theory. Validity of this description is further supported by
the fact that the SO(5) WZW theory can be regarded as a natural
extension of the SO(4) WZW theory for one
dimension,\cite{Tanaka_SO4} which nonabelian bosonization for spin
degrees of freedom or equivalently SU(2) chiral anomaly gives rise
to. This discovery is meaningful since it tells a possible
connection with the numerical simulation of Sandvik\cite{Sandvik}
although an appropriate treatment for the topological term is not
known, thus true solutions for such an effective field theory are
not found. Furthermore, such competition between
antiferromagnetism and valance bond solid is argued to be
reflected as the checker board pattern in the scanning tunnelling
microscopy.\cite{Sachdev_VBS_STM}

In this paper we follow the same strategy with the study of Tanaka
and Hu for spin degrees of freedom, but away from half filling
introducing charge fluctuations based on the SU(2) slave-boson
framework. Two important problems can arise away from half
filling. One cautious theorist may concern that a chemical
potential term will appear in the Dirac theory of the fermionic
spin representation away from half filling, thus breaking the
original structure for the SO(5) WZW theory. However, such a term
does not arise at least for the staggered flux phase of the SU(2)
slave-boson theory since spinons are still at half filling even
away from half filling. This is the reason why the SU(2)
slave-boson framework is utilized in the present study. The other
is how to construct couplings between doped holes and spin
fluctuations. The second problem is our key issue.

We find that it is not easy to derive such couplings if we resort
to the nonlinear $\sigma$ model description for the holon sector,
derived in the previous study.\cite{SU2_EFT} In the present paper
we perform fermionization for the holon sector via flux
attachment, where a Dirac theory with a chemical potential term
due to the presence of finite density of holons is found in the
staggered flux ansatz. Then, we can construct couplings between
spin fluctuations and holons in the same way as the spinon sector
since its theoretical structure is basically the same as that of
the spinon part. Performing the standard gradient expansion, we
have an effective field theory for SO(5) spin fluctuations,
fermionic holons and their couplings. As a result, dynamics of
doped holes in the antiferromagnetically correlated spin
background turns out to be described by the SO(5) WZW theory for
spin fluctuations, non-relativistic QED$_{3}$ around Dirac nodes
for doped holes, and interactions between valance bond
fluctuations and holes.

Fermionic holons remind us of the slave-fermion representation. We
compare our effective field theory with the U(1) slave-fermion
theory of the t-J model, and discuss similarities and differences
between them. Then, we solve the effective field theory in the
Eliashberg framework, where momentum dependence in the fermion
self-energy and vertex corrections are neglected, justified by
Migdal theorem and large $N$ approximation with the fermion flavor
number $N$ at least for the second assumption.\cite{FMQCP} The
Eliashberg framework allows us to construct Luttinger-Ward
functional of its closed form, thus making it possible to solve
the field theory self-consistently. Based on this framework, we
discuss its phase diagram, thermodynamics, and transport near and
away from the quantum critical point of our effective field
theory, where non-Fermi liquid physics is found, analogous with
that of the slave-fermion framework proposed
recently.\cite{ACL,NFL}

An important feature of our field theory is the coupling term
between valance bond fluctuations and doped holes. Considering
that valance bond fluctuations are deeply related with monopole
excitations of staggered U(1) gauge fields in the bosonic field
theory for spin fluctuations,\cite{Sachdev_NLsM} the presence of
such a coupling term allows us to investigate how dynamics of
doped holes affects deconfinement of bosonic spinons. Remarkably,
we find that such couplings help spin fractionalization near the
quantum critical point of the SO(5) WZW theory.

We also apply our field theory to one dimensional
antiferromagnetic doped Mott insulator, where spin fluctuations
are described by the SO(4) WZW theory while charge excitations are
represented by relativistic QED$_{2}$. Since one dimensional study
is well known, possible conclusions of our field theory in one
dimension can justify validity of our description. Furthermore,
reliable analytic techniques can be utilized in one dimension,
thus this study would suggest an important feature how doped holes
affect spinon deconfinement.

Finally, we discuss instability of the non-Fermi liquid metal
phase against superconductivity and density waves of doped holes.
As will be seen explicitly in the effective field theory [Eq.
(10)], interactions between doped holes turn out to be mediated by
both gauge and valance bond fluctuations. As a result, emergence
of superconductivity from the non-Fermi liquid state is determined
by competition between gauge and valance bond interactions. For
particle-hole channel instabilities, we discuss that the
homogeneous metallic phase can be stabilized at least when the
flavor number of fermions (doped holes) is sufficiently large.

\section{Effective field theory}

Dynamics of doped holes in the antiferromagnetically correlated
spin background is described by the t-J Hamiltonian \bqa && H = -
t \sum_{\langle i j \rangle} (c_{i\sigma}^{\dagger}c_{j\sigma} +
H.c.) + J \sum_{\langle i j \rangle} (\vec{S}_{i}\cdot\vec{S}_{j}
- \frac{1}{4}n_{i}n_{j}) . \nn \eqa Introducing an SU(2)
slave-boson representation for an electron field \bqa &&
c_{i\uparrow} = \frac{1}{\sqrt{2}} h_{i}^{\dagger} \psi_{i+} =
\frac{1}{\sqrt{2}} (b_{i1}^{\dagger}f_{i1} +
b_{i2}^{\dagger}f_{i2}^{\dagger}) , \nn && c_{i\downarrow} =
\frac{1}{\sqrt{2}} h_{i}^{\dagger} \psi_{i-} = \frac{1}{\sqrt{2}}
(b_{i1}^{\dagger}f_{i2} - b_{i2}^{\dagger}f_{i1}^{\dagger}) , \eqa
where $\psi_{i+} = \left(
\begin{array}{c} f_{i1} \\ f_{i2}^{\dagger}
\end{array}\right)$ and $\psi_{i-} =  \left(
\begin{array}{c} f_{i2} \\ - f_{i1}^{\dagger}
\end{array}\right)$ are SU(2)
spinon-spinors  and $h_{i} =  \left(
\begin{array}{c} b_{i1} \\ b_{i2}
\end{array}\right)$ is holon-spinor, one can rewrite the t-J model
in terms of these fractionalized excitations with hopping and
pairing fluctuations \bqa && L = L_{0} + L_{s} + L_{h} , ~~~~~
L_{0} = J_{r} \sum_{\langle i j \rangle}
\mathbf{tr}[U_{ij}^{\dagger}U_{ij}]   , \nn && L_{s} = \frac{1}{2}
\sum_{i} \psi_{i\alpha}^{\dagger} (\partial_{\tau} - i
a_{i0}^{k}\tau_{k})\psi_{i\alpha} \nn && + J_{r} \sum_{\langle i j
\rangle} ( \psi_{i\alpha}^{\dagger}U_{ij}\psi_{j\alpha} + H.c.) ,
\nn && L_{h} =  \sum_{i} h_{i}^{\dagger}(\partial_{\tau} - \mu  -
i a_{i0}^{k}\tau_{k})h_{i} \nn && + t_{r} \sum_{\langle i j
\rangle} ( h_{i}^{\dagger}U_{ij}h_{j} + H.c.)  , \eqa where the
SU(2) matrix field is $U_{ij} =
\left( \begin{array}{cc} - \chi_{ij}^{\dagger} & \eta_{ij} \\
\eta_{ij}^{\dagger} & \chi_{ij} \end{array}\right)$, and $J_{r} =
\frac{3J}{16}$ and $t_{r} = \frac{t}{2}$ are redefined
couplings.\cite{Lee_Nagaosa_Wen} Since this decomposition
representation enlarges the original electron Hilbert space,
constraints are introduced via Lagrange multiplier fields
$a_{i0}^{k}$ with $k = 1, 2, 3$.

In the SU(2) formulation Wen and Lee choose the staggered flux
gauge\cite{Lee_Nagaosa_Wen} \bqa && U_{ij}^{SF} = - \sqrt{\chi^{2}
+ \eta^{2}} \tau_{3} \exp[i(-1)^{i_{x} + i_{y}} \Phi \tau_{3}]
\eqa with a phase $\Phi = \tan^{-1}\Bigl( \frac{\eta}{\chi}
\Bigr)$. Although the staggered flux ansatz breaks translational
invariance, this formal symmetry breaking is restored via SU(2)
fluctuations between nearly degenerate U(1) mean-field states. For
example, one possible U(1) ground state, the d-wave pairing one
$U_{ij}^{dSC} = - \chi \tau_{3} + (-1)^{i_{y} + j_{y}} \eta
\tau_{1}$ can result from the SU(2) rotation $U_{ij}^{dSC}= W_{i}
U_{ij}^{SF} W_{j}^{\dagger}$ with an SU(2) matrix $W_{i} =
\exp\Bigl\{ i(-1)^{i_{x} + i_{y}}\frac{\pi}{4} \tau_{1} \Bigr\}$.
Then, our starting point becomes the following effective
Lagrangian \bqa && L_{SF} = \frac{1}{2} \sum_{i}
\psi_{i\alpha}^{\dagger} (\partial_{\tau} - i
a_{i0}^{3}\tau_{3})\psi_{i\alpha} \nn && + J_{r} \sum_{\langle i j
\rangle} ( \psi_{i\alpha}^{\dagger}U_{ij}^{SF}
e^{ia_{ij}^{3}\tau_{3}}\psi_{j\alpha} + H.c.) \nn && + \sum_{i}
h_{i}^{\dagger}(\partial_{\tau} - \mu  - i
a_{i0}^{3}\tau_{3})h_{i} \nn && + t_{r} \sum_{\langle i j \rangle}
( h_{i}^{\dagger}U_{ij}^{SF} e^{ia_{ij}^{3}\tau_{3}}h_{j} + H.c.)
\nn && + J_{r} \sum_{\langle i j \rangle}
\mathbf{tr}[U_{ij}^{SF\dagger}U_{ij}^{SF}] , \eqa where we have
introduced only one kind of gauge field $a_{\mu}^{3}$ as important
low energy fluctuations since other two ones, $a_{\mu}^{1}$ and
$a_{\mu}^{2}$ are gapped due to Anderson-Higgs mechanism in the
staggered flux phase.\cite{SU2_EFT}

Our idea is to fermionize the holon sector attaching a fictitious
flux to a holon field \bqa && L_{SF}^{h} = \sum_{i}
\eta_{i}^{\dagger}(\partial_{\tau} - \mu - i
a_{i0}^{3}\tau_{3})\eta_{i} \nn && + t_{r} \sum_{\langle i j
\rangle} ( \eta_{i}^{\dagger}U_{ij}^{SF} e^{ia_{ij}^{3}\tau_{3}}
e^{ic_{ij}\tau_{3}}\eta_{j} + H.c.) \nn && - i \sum_{i}
c_{i0}\Bigl( \eta_{i}^{\dagger} \tau_{3} \eta_{i} -
\frac{1}{2\Theta} (\partial_{x}c_{y} - \partial_{y}c_{x})_{i}
\Bigr) , \eqa where a bosonic field variable $h_{i}$ now becomes a
fermionic one $\eta_{i} = \left( \begin{array}{c} \eta_{i1} \\
\eta_{i2} \end{array}\right)$ with $\Theta = \pi$. It is important
to notice that our flux attachment is performed in an opposite way
for each isospin sector, confirmed by the presence of $\tau_{3}$
in $2 \Theta (\eta_{i}^{\dagger} \tau_{3} \eta_{i}) =
\partial_{x}c_{y} - \partial_{y}c_{x}$. As a result, there is no
net flux in the mean-field approximation of this construction,
considering that the density of $b_{i1}$ bosons is the same as
that of $b_{i2}$ bosons in the staggered flux phase.\cite{SU2_EFT}

This observation is interesting since it suggests a connection
with an SU(2) slave-fermion representation. If $a_{ij}^{3}$ is
shifted to $a_{ij}^{3} - c_{ij}$, the Chern-Simons flux is
transferred to spinons, turning their statistics into bosons.
Then, we have a bosonic spinon description with a fermionic holon,
nothing but the slave-fermion representation.

Performing the continuum approximation for the long wave length
and low energy limits, we find an effective field theory in terms
of only fermionic variables\cite{SU2_EFT} \bqa && {\cal L} =
\bar{\psi}\gamma_{\mu}(\partial_{\mu} - ia_{\mu}^{3}\tau_{3})\psi
+ \frac{1}{2e^{2}}
(\epsilon_{\mu\nu\gamma}\partial_{\nu}a_{\gamma}^{3})^{2} \nn && +
\bar{\eta}\gamma_{\mu}(\partial_{\mu} - ia_{\mu}^{3}\tau_{3} - i
c_{\mu}\tau_{3})\eta - \mu_{h}\bar{\eta}\gamma_{0}\eta \nn && +
\frac{i}{4\Theta}
c_{\mu}\epsilon_{\mu\nu\lambda}\partial_{\nu}c_{\lambda} . \eqa
Dirac structure\cite{Wen_Symmetry} results from the staggered flux
ansatz, where both $\psi$ and $\eta$ are $8$ component spinors and
Dirac gamma matrices are $\gamma_{0} = \left( \begin{array}{cc}
\sigma_{3} & 0
\\ 0 & - \sigma_{3} \end{array}\right)$, $\gamma_{1} = \left(
\begin{array}{cc} \sigma_{1} & 0 \\ 0 & - \sigma_{1}
\end{array}\right)$, and $\gamma_{2} = \left( \begin{array}{cc} \sigma_{2}
& 0 \\ 0 & - \sigma_{2} \end{array}\right)$. $a_{\mu}^{3}$ is
slave-boson U(1) gauge field, where a finite bare gauge charge $e$
is introduced. $c_{\mu}$ is Chern-Simons gauge field with its
statistical angle $\Theta = \pi$. It is important to understand
that spinons are still at half filling even away from half filling
in the SU(2) formulation. The single-occupancy constraint in the
SU(2) representation is given by $f_{i1}^{\dagger}f_{i1} +
f_{i2}^{\dagger}f_{i2} + b_{i1}^{\dagger}b_{i1} -
b_{i2}^{\dagger}b_{i2} = 1$. Thus, if the condition of $\langle
b_{i1}^{\dagger}b_{i1} \rangle = \langle b_{i2}^{\dagger}b_{i2}
\rangle = \frac{\delta}{2}$ with hole concentration $\delta$ is
satisfied, we see $\langle f_{i1}^{\dagger}f_{i1} +
f_{i2}^{\dagger}f_{i2} \rangle = 1$, i.e., spinons are at half
filling. As a result, a chemical potential term does not arise in
the spinon sector. Actually, this was demonstrated for the
staggered flux phase in the mean-field analysis of the SU(2)
slave-boson theory.\cite{SU2_EFT} On the other hand, a chemical
potential term appears in the holon sector to form a Fermi pocket
around the Dirac node.

To find an effective field theory for low energy spin
fluctuations, it is necessary to consider physical symmetry of the
spinon sector. We know that this symmetry is closely connected
with both spin and Dirac spaces. Since spin SU(2) symmetry is
hidden in the present $8$ component representation, we consider
the redundant representation $\Psi =  \left(
\begin{array}{c} \psi \\ \hat{\psi}
\end{array}\right)$ of $16$ components with a Dirac spinor $\hat{\psi} \equiv
i\tau_{2}\psi^{*} = \left(
\begin{array}{c} f_{\downarrow} \\ - f_{\uparrow}^{\dagger}
\end{array}\right)$, following Ref. \cite{Wen_Symmetry}.
Noting that the group space is composed of $G = G_{Dirac} \otimes
G_{gauge} \otimes G_{spin}$, we see $15$ generators associated
with SU(4) symmetry\cite{Wen_Symmetry} given by $I \otimes I
\otimes \vec{\sigma}$, $\gamma_{3} \otimes I \otimes
\vec{\sigma}$, $\gamma_{5} \otimes I \otimes \vec{\sigma}$,
$i\gamma_{3}\gamma_{5} \otimes I \otimes I$, $\gamma_{3}\otimes
\tau_{3} \otimes I$, $\gamma_{5} \otimes \tau_{3} \otimes I$, and
$i\gamma_{3}\gamma_{5} \otimes \tau_{3} \otimes \vec{\sigma}$,
where $\gamma_{3} = \left(
\begin{array}{cc} 0 & I \\ I & 0 \end{array}\right)$ and $\gamma_{5} =
\gamma_{0} \gamma_{1} \gamma_{2} \gamma_{3} = i \left(
\begin{array}{cc} 0 & I \\ - I & 0 \end{array}\right)$ satisfying
$[\gamma_{\mu},\gamma_{3(5)}]_{+} = 0$. This implies that symmetry
equivalent operators via SU(4) have the same strength for
instability, i.e., the same critical exponent for each correlation
function. Actually, this was intensively discussed in Refs.
\cite{ASL_Mother,Wen_Symmetry}.

Recently, C. Xu and S. Sachdev have claimed existence of a novel
spin liquid fixed point, where such an SU(4) symmetry is broken
down to SO(5).\cite{Sachdev_SO5} At this fixed point most relevant
spin fluctuations are Neel vector and valance bond fluctuations,
and they compete with each other since $10$ generators of $I
\otimes I \otimes \vec{\sigma}$, $\gamma_{3} \otimes I \otimes
\vec{\sigma}$, $\gamma_{5} \otimes I \otimes \vec{\sigma}$, and
$i\gamma_{3}\gamma_{5} \otimes I \otimes I$ remain and such spin
fluctuations are symmetry equivalent operators via chiral
rotation.

In this paper we also focus on competition between Neel and
valance-bond fluctuations. Following Tanaka and
Hu,\cite{Tanaka_SO5} we introduce an SO(5) superspin vector
$\vec{v} = (v_{1}, v_{2}, v_{3}, v_{4}, v_{5})$, where the former
three components form Neel vectors and the latter two ones
represent $x$ and $y$ valance bond fluctuations, and consider the
fermion mass term $- m \bar{\Psi} (\vec{v}\cdot\vec{\Gamma}) \Psi$
with $\vec{\Gamma} = (\sigma_{x}, \sigma_{y}, \sigma_{z}, i
\gamma_{3}, i \gamma_{5})$ for the spinon sector. On the other
hand, the holon mass term becomes $- m_{\eta} \bar{\eta}( i
\gamma_{3} v_{4} + i \gamma_{5} v_{5} )$ since it does not have
spin. As a result, we find the following Lagrangian for symmetry
"breaking" \bqa && {\cal L} =
\bar{\Psi}\gamma_{\mu}(\partial_{\mu} - ia_{\mu}^{3}\tau_{3})\Psi
- m \bar{\Psi} (\vec{v}\cdot\vec{\Gamma}) \Psi +
\frac{1}{2e^{2}}(\epsilon_{\mu\nu\gamma}\partial_{\nu}a_{\gamma}^{3})^{2}
\nn && + \bar{\eta}\gamma_{\mu}(\partial_{\mu} -
ia_{\mu}^{3}\tau_{3} - i c_{\mu}\tau_{3})\eta -
\mu_{h}\bar{\eta}\gamma_{0}\eta + \frac{i}{4\Theta}
c_{\mu}\epsilon_{\mu\nu\lambda}\partial_{\nu}c_{\lambda} \nn && -
m_{\eta} \bar{\eta}( i \gamma_{3} v_{4} + i \gamma_{5} v_{5} )
\eta . \eqa

The next task is to perform integration of Dirac spinons and
expand the resulting logarithmic action for the superspin vector.
Based on the gradient expansion method, Tanaka and Hu have derived
an SO(5) nonlinear $\sigma$ model with a WZW term ignoring gauge
fluctuations,\cite{Tanaka_SO5} $S_{spin} = S_{NLsM} + S_{WZW}$,
where $S_{NLsM} = \int {d^{3} x} \frac{1}{2g} (\partial_{\mu}
v_{k})^{2}$ and $S_{WZW} = i \frac{2\pi}{Area(S^{4})}
\int_{0}^{1}{dt} \int {d^{3} x} \epsilon_{abcde} v_{a}
\partial_{t} v_{b} \partial_{\tau} v_{c} \partial_{x} v_{d}
\partial_{y} v_{e}$ with $Area(S^{4}) =
\frac{2\pi^{5/2}}{\Gamma(5/2)}$. Since Dirac fermions are massive
in the symmetry "broken" phase and their fluctuations are ignored
in the low energy limit, gauge-fluctuation corrections will be
irrelevant considering that they can appear through fermion
bubbles. Only one point that should be careful is a topological
contribution, associated with an imaginary term. Although the WZW
term is nicely derived in the absence of gauge fluctuations, an
additional imaginary term may arise, a coupling term between a
topologically nontrivial fermionic current and gauge
field.\cite{Abanov_WZW} If we represent the Dirac spinor as
$\psi_{n} = \left( \begin{array}{c} \chi_{n}^{+} \\ \chi_{n}^{-}
\end{array}\right)$, where $\chi_{n}^{\pm}$ is a $2$ component
spinor with an isospin index $n = 1, 2$, one can see that each
sector in the Dirac space gives rise to such a term. However,
their signs are opposite, thus such terms are
cancelled.\cite{Kim_FNLsM} This is well known to be cancellation
of parity anomaly in condensed matter physics. Another way to say
this is that the signs of mass terms for Dirac fermions
($\chi_{n}^{+}$ and $\chi_{n}^{-}$) are opposite, resulting in
cancellation of the parity anomaly.

Based on the above discussion, we reach an effective field theory
\bqa && S = \int {d^{3} x} \Bigl[ \frac{1}{2g} (\partial_{\mu}
v_{k})^{2} - m_{\eta} \bar{\eta}( i \gamma_{3} v_{4} + i
\gamma_{5} v_{5} ) \eta \Bigr] + S_{WZW} \nn && + \int {d^{3} x}
\Bigl[ \bar{\eta}\gamma_{\mu}(\partial_{\mu} -
ia_{\mu}^{3}\tau_{3} - i c_{\mu}\tau_{3} - iA_{\mu})\eta -
\mu_{h}\bar{\eta}\gamma_{0}\eta \nn && + \frac{i}{4\Theta}
c_{\mu}\epsilon_{\mu\nu\lambda}\partial_{\nu}c_{\lambda} +
\frac{1}{2e^{2}}(\epsilon_{\mu\nu\gamma}\partial_{\nu}a_{\gamma}^{3})^{2}
\Bigr] , \eqa where spin fluctuations are described by the SO(5)
WZW theory. An interesting observation is that the Chern-Simons
contribution becomes irrelevant if the holon dynamics is in a
critical phase. Shifting the slave-boson gauge field as
$a_{\mu}^{3} - c_{\mu}$ and performing integration of Chern-Simons
gauge fields, we obtain $\sim (\partial\times\partial \times
a^{3})\cdot(\partial\times a^{3})$. This contribution is
irrelevant since it has a high scaling dimension owing to the
presence of an additional derivative. Considering that the density
of holons is finite to allow a Fermi surface (pocket around the
Dirac point), it is natural to assume that the fermion sector is
in criticality. We note that this kind of argument was well
utilized previously.\cite{Fisher_CS} As a result, we find an
effective field theory for an antiferromagnetic doped Mott
insulator problem \bqa && S = \int {d^{3} x} \Bigl[ \frac{1}{2g}
(\partial_{\mu} v_{k})^{2} - m_{\eta} \bar{\eta}( i \gamma_{3}
v_{4} + i \gamma_{5} v_{5} ) \eta \Bigr] \nn && + \int {d^{3} x}
\Bigl[ \bar{\eta}\gamma_{\mu}(\partial_{\mu} -
ia_{\mu}^{3}\tau_{3} - iA_{\mu})\eta -
\mu_{h}\bar{\eta}\gamma_{0}\eta \nn && +
\frac{1}{2e^{2}}(\epsilon_{\mu\nu\gamma}\partial_{\nu}a_{\gamma}^{3})^{2}
\Bigr] + S_{WZW} . \eqa

Several remarks are in order. First, the spin sector is described
by the SO(5) WZW theory even away from half filling, starting from
the SU(2) slave-boson theory in the staggered flux gauge. Validity
of this description will be further supported, comparing the
present effective theory with the slave-fermion framework and
applying it to one dimension. Second, dynamics of doped holes is
described by U(1) gauge theory with finite density of fermionic
holons around four Dirac nodes. Thus, non-Fermi liquid physics is
expected naturally. Third, interactions between spin fluctuations
and doped holes emerge as couplings between valance bond
fluctuations and fermionic holons. Considering that valance bond
fluctuations are deeply connected with monopole excitations of
CP$^{1}$ or staggered U(1) gauge fields in the bosonic field
theory for spin fluctuations, this coupling form implies how
dynamics of doped holes affects spin fractionalization, i.e.,
deconfinement of bosonic spinons. This will be discussed more
deeply.

\section{Connection with U(1) slave-fermion theory}

Although the effective field theory Eq. (10) has the similar
spirit with the slave-fermion framework, it has an important
different point. To understand this more clearly, we consider the
U(1) slave-fermion representation \bqa && c_{i\sigma} =
\psi_{i}^{\dagger}b_{i\sigma} , \eqa where $\psi_{i}$ and
$b_{i\sigma}$ are fermionic holon and bosonic spinon,
respectively. Inserting this representation into Eq. (1), we
obtain the following expression for both exchange-interaction and
electron-hopping terms, \bqa && J
\sum_{ij}(\vec{S}_{i}\cdot\vec{S}_{j} - \frac{1}{4}n_{i}n_{j}) \nn
&& \rightarrow \frac{J}{2}\sum_{ij}|\Delta_{ij}^{b}|^{2} -
J\sum_{ij}(\Delta_{ij}^{b\dagger}\epsilon_{\alpha\beta}b_{i\alpha}b_{j\beta}
+ H.c.)  , \nn && - t \sum_{ij}(c_{i\sigma}^{\dagger}c_{j\sigma} +
H.c.) \rightarrow   t\sum_{ij}(\chi_{ji}^{\psi}\chi_{ij}^{b} +
H.c.) \nn && - t
\sum_{ij}(b_{i\sigma}^{\dagger}\chi_{ij}^{b}b_{j\sigma} + H.c.) +
t \sum_{ij}( \psi_{j}^{\dagger}\chi_{ji}^{\psi}\psi_{i} + H.c.) ,
\nn \eqa where each composite field is given by \bqa \Delta_{ij} =
\sum_{\alpha\beta} \epsilon_{\alpha\beta}b_{i\alpha}b_{j\beta} ,
~~~ \chi_{ji}^{\psi} = \sum_{\sigma} b_{i\sigma}^{\dagger}
b_{j\sigma} , ~~~ \chi_{ij}^{b} = \psi_i \psi_j^{\dagger} ,
\nonumber \eqa representing short-range antiferromagnetic
correlations, ferromagnetic spin fluctuations, and hopping of
holons, respectively.

In the low energy and long wave-length limits one can set the
above collective fields as \bqa && \Delta_{ij} = \Delta
e^{ic_{ij}} , ~~~ \chi_{ij}^{b} = \chi_{b} e^{ia_{ij}} , ~~~
\chi_{ij}^{\psi} = \chi_{\psi} e^{ia_{ij}} , \eqa where amplitude
fluctuations are frozen to be their saddle-point values, and only
phase fluctuations are kept importantly. Then, the resulting
slave-fermion Lagrangian becomes \bqa && L_{eff} = L_{s} + L_{h} +
L_{c} + L_{0} , \nn && L_{s} =
\sum_{i}b_{i\sigma}^{\dagger}(\partial_{\tau} - \mu)b_{i\sigma} -
t \chi_{b} \sum_{ij}(b_{i\sigma}^{\dagger}e^{ia_{ij}}b_{j\sigma} +
H.c.) \nn && - J \Delta \sum_{ij}(e^{-
ic_{ij}}\epsilon_{\alpha\beta}b_{i\alpha}b_{j\beta} + H.c.) , \nn
&& L_{h} = \sum_{i}\psi_{i}^{\dagger}\partial_{\tau}\psi_{i} + t
\chi_{\psi} \sum_{ij}( \psi_{j}^{\dagger}e^{- ia_{ij}}\psi_{i} +
H.c.) , \nn && L_{c} = i
\sum_{i}\lambda_{i}(b_{i\sigma}^{\dagger}b_{i\sigma} +
\psi_{i}^{\dagger}\psi_{i} - 2S)  , \nn && L_{0} = N_{L}(
J\Delta^{2} + 4t \chi_{\psi}\chi_{b}) , \eqa where $\lambda_{i}$
in $L_{c}$ is a Lagrange multiplier field to impose the single
occupancy constraint with $S = 1/2$, and $N_{L}$ in $L_{0}$ is
number of lattice sites.

Comparing Eq. (10) with Eq. (14), one will find several different
points. First, the spectrum of holon excitations is not
relativistic in Eq. (14). However, this can be adjusted, allowing
$\pi$ flux in the U(1) gauge field $a_{ij}$, i.e., $\sum_{\Box}
a_{ij} = \pi$. Then, four Dirac nodes arise, and hole doping gives
finite density of holons, resulting in the hole pockets around the
nodes.

Second, spin dynamics is described by usual spin $1$ excitations
in Eq. (10) while it is expressed with spin $1/2$ fractionalized
spinons in Eq. (14). However, the presence of the WZW term in Eq.
(10) may give rise to spin fractionalization as one dimensional
physics\cite{Tanaka_SO4} or the previous
proposal\cite{Senthil_DQCP} for deconfined quantum criticality. If
we start from Eq. (14), we focus on two kinds of gauge
fluctuations, corresponding to staggered U(1) gauge fields
$c_{ij}$ and uniform U(1) gauge fields $a_{ij}$, respectively. In
particular, the staggered U(1) gauge field $c_{ij}$ turns out to
be the same as the CP$^{1}$ gauge field if the Schwinger-boson
effective Lagrangian is mapped onto the CP$^{1}$ gauge theory of
the O(3) nonlinear $\sigma$ model in the long wave-length
limit.\cite{Sachdev_NLsM} In this respect, if we assume that
staggered gauge fluctuations mediate confining interactions
between spinons, we expect to see spin $1$ fluctuations described
by a $\sigma$ model-type theory. However, the fate of valance bond
fluctuations is not clear in this case.

Third, the coupling structure between spin fluctuations and doped
holes in Eq. (10) differs from that in Eq. (14). In the
slave-fermion Lagrangian Eq. (14) direct couplings between spin
and charge degrees of freedom do not exist although nonlocal
current-current interactions appear in the low energy limit,
associated with the single occupancy constraint. Even if we assume
staggered gauge fluctuations confining, it is difficult to find a
coupling term between spin fluctuations and doped holes. On the
other hand, our effective field theory Eq. (10) exhibits its
direct coupling term explicitly. This coupling term is an
important ingredient for dynamics of doped holes in the quantum
antiferromagnet. As demonstrated intensively in the previous
work,\cite{Sachdev_NLsM} valance bond fluctuations are deeply
related with instanton (monopole) excitations of staggered U(1)
gauge fields $c_{ij}$. Actually, $v_{4}$ and $v_{5}$ fields can be
identified with an instanton operator in the presence of a
topological $\theta$ term. This implies that the presence of doped
holes affects spinon confinement directly via their couplings. In
the cumulant expansion for the coupling term one can see that
holon dynamics modifies monopole-monopole correlations, giving
rise to dissipation. As a result, an effective theory for monopole
excitations away from half filling is changed seriously from that
at half filling. In the next section we will see that the presence
of doped holes helps bosonic spinons deconfined near the quantum
critical point.

\section{Physical properties}

\subsection{Eliashberg framework}

Three kinds of field variables, that is, fermionic holon, SO(5)
superspin vector, and uniform U(1) gauge field make our effective
field theory complicated. In this respect it is not easy to treat
such all degrees of freedom self-consistently. Recently, it was
explicitly demonstrated that Eliashberg framework is the minimal
self-consistent treatment for an effective field theory near its
quantum critical point.\cite{FMQCP} The Eliashberg treatment
neglects momentum dependence of a fermion self-energy and vertex
corrections. The first assumption is based on the fact that
momentum dependence of a fermion self-energy is regular, and
singular physics arises from its frequency dependence. This can be
checked explicitly at least in the one loop level. The second
assumption is more serious than the first one, sometimes called
Migdal theorem.\cite{Migdal} When fermions are much faster than
bosons, vertex corrections can be neglected since the pre-factor
in the renormalized vertex is given by the ratio of fermion and
boson velocities. However, this turns out to be not sufficient for
the Eliashberg framework.\cite{FMQCP} Another parameter is shown
to be need, that is, the fermion flavor number $N$. In the large
$N$ limit the Eliashberg framework is justified.

For the Eliashberg treatment we rewrite the effective field theory
as follows \bqa && S = S_{\eta} + S_{v} + S_{a} + S_{int} , \nn &&
S_{\eta} = \int{d^3x} \Bigl\{
\bar{\eta}\gamma_{\mu}(\partial_{\mu} - iA_{\mu})\eta -
\mu_{h}\bar{\eta}\gamma_{0}\eta \Bigr\} , \nn && S_{v} =
\int{d^{3}x} \Bigl\{ \frac{1}{2g} (\partial_{\mu} v_{k})^{2} +
m_{v}^{2} (|v_{k}|^{2} - 1) \Bigr\} + S_{WZW} , \nn && S_{a} =
\int{d^{3}x} \Bigl\{
\frac{1}{2e^{2}}(\epsilon_{\mu\nu\gamma}\partial_{\nu}a_{\gamma}^{3})^{2}
\Bigr\} , \nn &&  S_{int} = \int{d^{3}x} \Bigl\{- m_{\eta}
\bar{\eta}( i \gamma_{3} v_{4} + i \gamma_{5} v_{5} )\eta - i
a_{\mu}^{3}\bar{\eta}\gamma_{\mu} \tau_{3} \eta \Bigr\} , \nn \eqa
where $m_{v}^{2}$ represents mass of superspin vector bosons
arising from the SO(5) rotor constraint $\sum_{k=1}^{5}|v_{k}|^{2}
=1$. It will be determined self-consistently.

Performing the cumulant expansion for $S_{int}$, we find an
effective action $S_{eff} = S_{v} + S_{\eta} + S_{a} - \frac{1}{2}
\Bigl(\langle S_{int}^{2} \rangle - \langle S_{int} \rangle^{2}
\Bigr)$. Although this expression is for the second order, it
includes an infinite order actually. Accordingly, we can construct
the corresponding Luttinger-Ward functional in the Eliashberg
framework \bqa && F_{LW} = F_{LW}^{\eta} + F_{LW}^{v} + F_{LW}^{a}
+ Y_{v} + Y_{a} , \nn && F_{LW}^{\eta} = - T \sum_{i\omega}
\int\frac{d^{d}k}{(2\pi)^{d}} \mathbf{tr} \Bigl[\ln\Bigl\{
g_{\eta}^{-1}(k,i\omega) + \Sigma_{\eta}(i\omega) \Bigr\} \nn && -
\Sigma_{\eta}(i\omega) G_{\eta}(k,i\omega) \Bigr] , \nn &&
F_{LW}^{v} = T \sum_{i\Omega} \int\frac{d^{d}q}{(2\pi)^{d}} \Bigl[
\sum_{m,n=1}^{5} \ln\Bigl\{ d_{v}^{-1}(q,i\Omega)\delta_{mn} \nn
&& + \Pi_{v}^{mn}(q,i\Omega) \Bigr\} - \sum_{m,n=1}^{5}
\Pi_{v}^{mn}(q,i\Omega)D_{v}^{mn}(q,i\Omega)\Bigr] - m_{v}^{2} ,
\nn && F_{LW}^{a} = T \sum_{i\Omega} \int\frac{d^{d}q}{(2\pi)^{d}}
\Bigl[\ln\Bigl\{ d_{a}^{-1}(q,i\Omega) + \Pi_{a}(q,i\Omega)
\Bigr\} \nn && - \Pi_{a}(q,i\Omega) D_{a}(q,i\Omega) \Bigr] , \nn
&& Y_{v} = - \frac{m_{\eta}^{2}}{2} T \sum_{i\omega}
\int\frac{d^{d}k}{(2\pi)^{d}} T \sum_{i\Omega}
\int\frac{d^{d}q}{(2\pi)^{d}} \nn && \sum_{m,n=4}^{5} \mathbf{tr}[
D_{v}^{mn}(q,i\Omega) \gamma_{m} G_{\eta}(k+q,i\omega+i\Omega)
\gamma_{n}G_{\eta}(k,i\omega)] , \nn && Y_{a} = - \frac{1}{2} T
\sum_{i\omega} \int\frac{d^{d}k}{(2\pi)^{d}} T \sum_{i\Omega}
\int\frac{d^{d}q}{(2\pi)^{d}} \nn && \mathbf{tr} [
D_{a}^{\mu\nu}(q,i\Omega) \gamma_{\mu}\tau_{3}
G_{\eta}(k+q,i\omega+i\Omega)
\gamma_{\nu}\tau_{3}G_{\eta}(k,i\omega)]  . \eqa
$G_{\eta}(k,i\omega)$ in $F_{LW}^{\eta}$ is a renormalized
propagator for holons, given by $G_{\eta}(k,i\omega) =
\Bigl\{g_{\eta}^{-1}(k,i\omega) + \Sigma_{\eta}(i\omega)
\Bigr\}^{-1}$, where $g_{\eta}(k,i\omega) = \Bigl( i \gamma_{0}
\omega + i \gamma_{i} k_{i} + \mu_{h} \gamma_{0} \Bigr)^{-1}$ is
its bare propagator, and $\Sigma_{\eta}(i\omega)$ is its
momentum-independent self-energy. $D_{v}^{mn}(q,i\Omega)$ in
$F_{LW}^{v}$ is a renormalized propagator for superspin vector
fields, given by $D_{v}^{mn}(q,i\Omega) =
\Bigl\{d_{v}^{-1}(q,i\Omega)\delta_{mn} +
\Pi_{v}^{mn}(q,i\Omega)\Bigr\}^{-1}$, where $d_{v}(q,i\Omega) =
\Bigl( \frac{q^{2} + \Omega^{2}}{2g} + m_{v}^{2} \Bigr)^{-1}$ is
its bare propagator, and $\Pi_{v}^{mn}(q,i\Omega)$ is its
self-energy. $D_{a}(q,i\Omega)$ in $F_{LW}^{a}$ is a renormalized
kernel for the gauge propagator $D_{a}^{\mu\nu}(q,i\Omega) =
D_{a}(q,i\Omega)\Bigl( \delta_{\mu\nu} -
\frac{q_{\mu}q_{\nu}}{q^{2}} \Bigr)$, given by $D_{a} (q,i\Omega)
= \Bigl\{d_{a}^{-1}(q,i\Omega) + \Pi_{a}(q,i\Omega) \Bigr\}^{-1}$,
where $d_{a}(q,i\Omega) = \Bigl( \frac{q^{2}+\Omega^{2}}{2e^{2}}
\Bigr)^{-1}$ is its bare kernel, and $\Pi_{a}(q,i\Omega)$ is its
self-energy in $\Pi_{a}^{\mu\nu}(q,i\Omega) =
\Pi_{a}(q,i\Omega)\Bigl( \delta_{\mu\nu} -
\frac{q_{\mu}q_{\nu}}{q^{2}} \Bigr)$. $Y_{v}$ is introduced for
self-energy corrections resulting from the first term in $S_{int}$
of Eq. (15) while $Y_{a}$ is for those arising from the second
term of $S_{int}$. In the self-energy functional $Y_{v}$
$\gamma_{4}$ should be replaced with $\gamma_{3}$, where this
problem appears from our notation form.

It is important to notice that Luttinger-Ward functional is not
usually written in a closed form. However, the Luttinger-Ward
functional can be written in its closed form at least for the
Eliashberg framework.\cite{Chubukov_FL} Actually, performing
variation for the Luttinger-Ward functional with respect to each
self-energy, i.e., $\frac{\delta F_{LW}}{\delta
\Sigma_{\eta}(i\omega)} = 0$, $\frac{\delta F_{LW}}{\delta
\Pi_{v}^{mn}(q,i\Omega)} = 0$, and $\frac{\delta F_{LW}}{\delta
\Pi_{a}^{\mu\nu}(q,i\Omega)} = 0$, we find self-consistent
Eliashberg equations \bqa && \Sigma_{\eta}(i\omega) = m_{\eta}^{2}
T \sum_{i\Omega} \int \frac{d^{d}q}{(2\pi)^{d}} \sum_{m,n=4}^{5}
\nn && D_{v}^{mn}(q,i\Omega) \gamma_{m}
G_{\eta}(k_{F}+q,i\omega+i\Omega) \gamma_{n} + T \sum_{i\Omega}
\int \frac{d^{d}q}{(2\pi)^{d}} \nn && D_{a}^{\mu\nu}(q,i\Omega)
\gamma_{\mu}\tau_{3} G_{\eta}(k_{F}+q,i\omega+i\Omega)
\gamma_{\nu}\tau_{3} , \nn && \Pi_{v}^{mn}(q,i\Omega) = T
\sum_{i\omega}\int \frac{d^{d}k}{(2\pi)^{d}} \sum_{m,n=4}^{5} \nn
&& \Bigl( - \frac{m_{\eta}^{2}}{2} \mathbf{tr}
[\gamma_{m}G_{\eta}(k+q,i\omega+i\Omega)\gamma_{n}G_{\eta}(k,i\omega)]
\Bigr) , \nn && \Pi_{a}^{\mu\nu}(q,i\Omega) = T \sum_{i\omega}
\int\frac{d^{d}k}{(2\pi)^{d}} \nn && \Bigl( - \frac{1}{2}
\mathbf{tr} [ \gamma_{\mu}\tau_{3} G_{\eta}(k+q,i\omega+i\Omega)
\gamma_{\nu}\tau_{3}G_{\eta}(k,i\omega)] \Bigr) . \eqa The holon
self-energy results from both valance bond and gauge fluctuations,
where $\gamma_{4} \rightarrow \gamma_{3}$ is performed. The
superspin vector self-energy arises from holon fluctuations, where
$\gamma_{4}$ is also replaced with $\gamma_{3}$. Notice
$\Pi_{v}^{mn}(q,i\Omega) = 0$ for $m, n = 1, 2, 3$. The gauge
field self-energy appears from holon current fluctuations. Eqs.
(16) and (17) complete the Eliashberg framework.

\subsection{Simplification of Luttinger-Ward functional}

Using the Eliashberg equations (17), one can simplify the
Luttinger-Ward functional Eq. (16) as follows \bqa && F_{LW} = - T
\sum_{i\omega} \int\frac{d^{d}k}{(2\pi)^{d}} \mathbf{tr}
\Bigl[\ln\Bigl\{ g_{\eta}^{-1}(k,i\omega) + \Sigma_{\eta}(i\omega)
\Bigr\} \nn && - \Sigma_{\eta}(i\omega) G_{\eta}(k,i\omega) \Bigr]
\nn && + T \sum_{i\Omega} \int\frac{d^{d}q}{(2\pi)^{d}} \Bigl[
\sum_{m,n=1}^{5} \ln\Bigl\{ d_{v}^{-1}(q,i\Omega)\delta_{mn} +
\Pi_{v}^{mn}(q,i\Omega) \Bigr\} \Bigr] \nn && - m_{v}^{2} + T
\sum_{i\Omega} \int\frac{d^{d}q}{(2\pi)^{d}} \ln\Bigl\{
d_{a}^{-1}(q,i\Omega) + \Pi_{a}(q,i\Omega) \Bigr\} , \eqa where
the self-energy parts for superspin vector and U(1) gauge fields
are cancelled out from the use of Eq. (17). Then, one can see that
the holon free energy is nothing but the free energy of Fermi
liquid as follows \bqa && F_{LW}^{\eta} \approx - N_{\eta} T
\sum_{i\omega} \int\frac{d^{d}k}{(2\pi)^{d}} \ln\Bigl\{ 2
\mu_{h}(i\omega) + \mu_{h}^{2} - \omega^{2} - k^{2} \Bigr\} \nn &&
\approx - N_{\eta} \rho_{\eta} T \sum_{i\omega} |\omega| = -
\frac{\pi N_{\eta} \rho_{\eta}}{6} T^{2} = F_{FL}^{\eta} . \eqa
Here, $\rho_{\eta}$ is the density of states around the Dirac
node, and the number of Dirac nodes is $N_{\eta} = 4$. $f(y) =
\frac{1}{e^{y} + 1}$ is the Fermi-Dirac distribution function with
temperature scaling. As a result, we find the Eliashberg free
energy of our effective field theory \bqa && F_{LW} = - \frac{\pi
N_{\eta} \rho_{\eta}}{6} T^{2} - m_{v}^{2} \nn && + T
\sum_{i\Omega} \int\frac{d^{d}q}{(2\pi)^{d}} \Bigl[
\sum_{m,n=1}^{5} \ln\Bigl\{ d_{v}^{-1}(q,i\Omega)\delta_{mn} +
\Pi_{v}^{mn}(q,i\Omega) \Bigr\} \Bigr] \nn && + T \sum_{i\Omega}
\int\frac{d^{d}q}{(2\pi)^{d}} \ln\Bigl\{ d_{a}^{-1}(q,i\Omega) +
\Pi_{a}(q,i\Omega) \Bigr\} . \eqa

The remaining thing is to evaluate each self-energy. The superspin
vector self-energy $\Pi_{v}^{mn}(q,i\Omega) = \Pi_{v}(q,i\Omega)
\delta_{mn}$ for $m, n = 4, 5$ is found to be \bqa &&
\Pi_{v}(q,i\Omega) \approx \frac{\pi N_{\eta}
m_{\eta}^{2}\rho_{\eta}}{4} \frac{|\Omega|}{q} . \eqa This is
nothing but the standard Landau damping term, originating from
fermion excitations near the Fermi surface. Since the density of
holons is finite due to hole doping, emergence of the Landau
damping term is quite natural. Then, the renormalized propagator
for superspin fluctuations is given by \bqa &&
D_{v}^{mn}(q,i\Omega) = \frac{\delta_{mn}}{ \frac{q^{2} +
\Omega^{2}}{2g} + m_{v}^{2} } , ~~~~~ \mbox{for} ~~~ m, n = 1, 2,
3 , \nn && D_{v}^{mn}(q,i\Omega) = \frac{\delta_{mn}}{ \frac{q^{2}
+ \Omega^{2}}{2g} + m_{v}^{2} + \frac{\pi N_{\eta}
m_{\eta}^{2}\rho_{\eta}}{4} \frac{|\Omega|}{q} } \nn && \approx
\frac{\delta_{mn}}{ \frac{q^{2}}{2g} + m_{v}^{2} + \frac{\pi
N_{\eta} m_{\eta}^{2}\rho_{\eta}}{4} \frac{|\Omega|}{q} }  , ~~~~~
\mbox{for} ~~~ m, n = 4, 5 , \eqa implying that antiferromagnetic
spin fluctuations are described by $z = 1$ theory while valance
bond fluctuations are expressed by $z = 3$ theory, where $z$ is
the dynamical exponent. The gauge self-energy is also given by the
Landau damping term $\Pi_{a}(q,i\Omega) = \frac{\pi N_{\eta}
\rho_{\eta}}{4} \frac{|\Omega|}{q}$, thus the renormalized gauge
propagator becomes \bqa && D_{a}(q,i\Omega) \approx
\frac{1}{\frac{q^{2}}{2e^{2}} + \frac{\pi N_{\eta} \rho_{\eta}}{4}
\frac{|\Omega|}{q}} , \eqa described by $z = 3$ critical theory.

Inserting bosonic self-energies into Eq. (20), we find the final
expression for the Eliashberg free energy \bqa && F_{LW} = -
\frac{\pi N_{\eta} \rho_{\eta}}{6} T^{2} - \frac{\xi^{-2}}{2g} + T
\sum_{i\Omega} \int\frac{d^{d}q}{(2\pi)^{d}} \nn && \Bigl\{ 3
\ln\Bigl(q^{2} + \Omega^{2} + \xi^{-2}\Bigr) + 2 \ln \Bigl( q^{2}
+ \xi^{-2} + \gamma_{v} \frac{|\Omega|}{q} \Bigr) \Bigr\} \nn && +
T \sum_{i\Omega} \int\frac{d^{d}q}{(2\pi)^{d}} \ln\Bigl( q^{2} +
\gamma_{a} \frac{|\Omega|}{q} \Bigr) , \eqa where \bqa && \xi^{-2}
= 2g m_{v}^{2} , ~~~ \gamma_{v} = \frac{ \pi g N_{\eta}
m_{\eta}^{2}\rho_{\eta}}{2} , ~~~ \gamma_{a} = \frac{ \pi e^{2}
N_{\eta} \rho_{\eta}}{2} \nonumber \eqa represent the correlation
length for superspin fluctuations, Landau damping coefficient for
superspin fields, and that for gauge fields.

Several remarks about the Eliashberg framework are in order.
First, the Landau damping contribution does not depend on the
fermion self-energy in the Eliashberg framework, allowing one to
use a bare fermion propagator instead of its full green's
function.\cite{FMQCP} Second, $z = 3$ criticality supports Migdal
theorem since on-shell (or resonance) fermion momenta are larger
than on-shell boson one at the same energy. Third, the number of
Dirac nodes $N_{\eta}$ plays the same role as the fermion flavor
number in the Eliashberg framework. In this respect the Eliashberg
framework works well for our effective field theory.

\subsection{Phase diagram}

Performing variation of the free energy Eq. (24) with respect to
the correlation length, i.e., $\frac{\partial F_{LW}}{\partial
\xi^{-2}} = 0$, we obtain the self-consistent equation for the
correlation length in the Eliashberg framework \bqa && 1 = 2g T
\sum_{i\Omega} \int\frac{d^{d}q}{(2\pi)^{d}} \Bigl( \frac{3 }{
q^{2} + \Omega^{2} + \xi^{-2} } \nn && + \frac{2}{ q^{2} +
\xi^{-2} + \gamma_{v} \frac{|\Omega|}{q} }\Bigr) . \eqa Notice
that interactions between valance bond fluctuations and holons
result in the $z=3$ part.

Performing integration, we find the following expression for the
correlation length \bqa && 1 = \frac{3g}{\pi} T \Bigl\{ \ln
\sinh\Bigl( \frac{\Lambda}{2T} \Bigr) - \ln \sinh\Bigl(
\frac{\xi^{-1}}{2T} \Bigr) \Bigr\} \nn && + \frac{4g }{3\pi} T
\ln\Bigl\{\sinh\Bigl(\frac{(\xi^{3}\gamma_{v})^{-1}}{2T}\Bigr)
\Bigr\} \nn && - 2T \Bigl[ - \frac{(\xi^{3}\gamma_{v})^{-2}}{4T} -
\frac{\pi^{2}}{6} T - (\xi^{3}\gamma_{v})^{-1} \ln \Bigl( 1 -
e^{-\frac{(\xi^{3}\gamma_{v})^{-1}}{T}} \Bigr) \nn && +
(\xi^{3}\gamma_{v})^{-1} \ln \Bigl\{ \sinh\Bigl(
\frac{(\xi^{3}\gamma_{v})^{-1}}{2T} \Bigr) \Bigr\}  + T
\sum_{k=1}^{\infty} \frac{e^{- k
\frac{(\xi^{3}\gamma_{v})^{-1}}{T}}}{k^{2}} \Bigr] \nn && +
\frac{2g}{3} T \Bigl[ \ln \sinh\Bigl( \frac{\Lambda}{2T} \Bigr) -
\ln \sinh\Bigl( \frac{(\xi^{3}\gamma_{v})^{-1}}{2T} \Bigr) \Bigr]
. \eqa As shown in this equation, there exist three regimes, (A)
$T < \frac{(\xi^{3}\gamma_{v})^{-1}}{2} < \frac{\xi^{-1}}{2}$, (B)
$\frac{(\xi^{3}\gamma_{v})^{-1}}{2} < T < \frac{\xi^{-1}}{2}$, and
(C) $\frac{(\xi^{3}\gamma_{v})^{-1}}{2} < \frac{\xi^{-1}}{2} < T$,
emerging from coexistence of $z = 1$ (antiferromagnetic) and $z =
3$ (valance bond) fluctuations. In regime (A) both $z = 1$ and $z
= 3$ fluctuations are gapped while in regime (C) both spin
fluctuations are critical, that is, in the quantum critical
regime. In regime (B) only valance bond fluctuations ($z = 3$) are
critical, and $z = 1$ antiferromagnetic ones are gapped. The phase
diagram is shown in Fig. 1.

\begin{figure}[t]
\vspace{5cm} \includegraphics{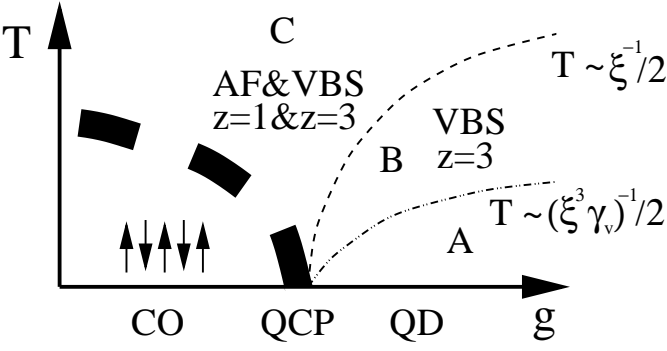} \caption{ (Color online)
Schematic phase diagram : CO, QD, and QCP represent classically
ordered, quantum disordered, and quantum critical point,
respectively. Two crossover scales emerge as $T \sim
\frac{(\xi^{3}\gamma_{v})^{-1}}{2}$ and $T \sim
\frac{\xi^{-1}}{2}$, identifying three regimes. In regime (A) of
$T < \frac{(\xi^{3}\gamma_{v})^{-1}}{2} < \frac{\xi^{-1}}{2}$ both
$z = 1$ AF and $z = 3$ VBS fluctuations are gapped while in regime
(C) of $\frac{(\xi^{3}\gamma_{v})^{-1}}{2} < \frac{\xi^{-1}}{2} <
T$ both spin fluctuations are critical, that is, in the quantum
critical regime. In regime (B) of
$\frac{(\xi^{3}\gamma_{v})^{-1}}{2} < T < \frac{\xi^{-1}}{2}$ only
$z = 3$ VBS fluctuations are critical, and $z = 1$ AF ones are
gapped.} \label{fig1}
\end{figure}

Considering $\frac{(\xi^{3}\gamma_{v})^{-1}}{2} \ll
\frac{\xi^{-1}}{2}$ near the quantum critical point, we can
simplify the above equation, and find that antiferromagnetic
fluctuations play an important role in determining the SO(5)
superspin correlation length near the quantum critical point. As a
result, we obtain the following expression for the correlation
length \bqa \xi^{-1}(g \sim g_{c}; T < \frac{\xi^{-1}}{2}) &&
\approx \Bigl\{ \Bigl( 1 + \frac{2\pi}{9} \Bigr) \Lambda -
\frac{2\pi}{3g} \Bigr\} \nn && \equiv \frac{2\pi}{3} \Bigl(
\frac{1}{g_{c}} - \frac{1}{g} \Bigr) , \nn \xi^{-1} (g \sim g_{c};
T > \frac{\xi^{-1}}{2}) && = 2 T e^{\frac{\pi}{3} \frac{\Bigl(
\frac{1}{g_{c}} - \frac{1}{g} \Bigr)}{T}} , \eqa where $g_{c} =
\Bigl( \frac{3}{2\pi} + \frac{1}{3} \Bigr) \Lambda$ is the quantum
critical point.\cite{SO5} Recalling $g(\delta) \propto
m^{2}(\delta)$ in the gradient expansion and $m^{2}(\delta)
\propto |\delta - \delta_{0}|$ with hole concentration $\delta <
\delta_{0}$ near the quantum critical point, we find the critical
hole concentration in the Eliashberg framework \bqa && \delta_{c}
= \delta_{0} - \frac{1}{c} \Bigl\{\Bigl( \frac{3}{2\pi} +
\frac{1}{3} \Bigr) \Lambda\Bigr\} , \eqa where $c$ is a positive
numerical constant. Thus, an antiferromagnetically ordered phase
appears in $\delta < \delta_{c}$, and a quantum disordered state
arises in $\delta > \delta_{c}$. The nature of the disordered
state will be determined by SO(5) symmetry breaking effective
potentials.

\subsection{Deconfined quantum critical point}

We consider how interactions between valance bond fluctuations and
holons affect the deconfined quantum critical point, proposed to
exist at half filling. As discussed before, valance bond
fluctuations can be identified with monopole excitations of
staggered gauge fields. Introducing $\Psi = \frac{1}{\sqrt{2}}
(v_{4} + i v_{5})$ and considering SO(5) symmetry breaking for the
WZW term as follows in Ref. \cite{Tanaka_SO5}, we find an
effective field theory for monopole or valance bond fluctuations
from the effective field theory Eq. (10), \bqa && S_{VB} = T
\sum_{i\Omega} \int \frac{d^{2}q}{(2\pi)^{2}} \nn &&
\Psi^{\dagger}(q,i\Omega) \Bigl( \frac{q^{2}}{2g} + \frac{\pi
N_{\eta} m_{\eta}^{2}\rho_{\eta}}{4} \frac{|\Omega|}{q} \Bigr)
\Psi(q,i\Omega) \nn && - \int_{0}^{\beta}{d\tau} \int d^{2} r
y_{m} (\Psi^{4} + \Psi^{\dagger 4}) . \eqa Here, the cubic power
in the last term results from the WZW term with SO(5) symmetry
breaking, where $y_{m}$ is the monopole fugacity. If the
topological $\theta$ term is not taken into account, the
condensation-induced term will be given by $-
\int_{0}^{\beta}{d\tau} \int d^{2} r y_{m} (\Psi +
\Psi^{\dagger})$.

An important point is that dynamics of valance bond excitations is
described by $z = 3$ critical theory at the quantum critical
point. As a result, two spacial dimensions are already in the
upper critical dimension, thus higher order interactions beyond
the Gaussian term are irrelevant. It means that the WZW-induced
cubic term can be neglected safely at the quantum critical point.
Equivalently, the monopole fugacity vanishes at the quantum
critical point, indicating deconfinement of bosonic spinons. We
point out that the topological term plays an important role for
deconfinement even away from half filling. If the topological term
is ignored, the monopole-fugacity term of linear in $\Psi$ is
relevant, giving rise to confinement.

Although proliferation of monopole excitations is prohibited due
to fermion excitations, monopoles are in a critical state of $z =
3$ instead of a gapped phase. Such critical valance bond
fluctuations will affect holon dynamics. This feedback effect to
holon dynamics is an important ingredient for the source of
non-Fermi liquid physics near the quantum critical point.

\subsection{Thermodynamics}

We study thermodynamics in each regime. We start from the
following expression for the Eliashberg free energy \bqa && F_{LW}
=  - \frac{\pi N_{\eta} \rho_{\eta}}{6} T^{2} -
\frac{\xi^{-2}}{2g} \nn && + \frac{3}{2\pi} T \int_{0}^{\infty} d
q q \ln \Bigl\{ 2 \sinh \Bigl( \frac{\sqrt{q^{2} + \xi^{-2}}}{2T}
\Bigr) \Bigr\} \nn && - \frac{1}{\pi^{2}} \int_{0}^{\infty} d q q
\int_{0}^{\infty} {d \nu} \coth\Bigl(\frac{\nu}{2T}\Bigr)
\tan^{-1}\Bigl( \gamma_{v} \frac{\nu}{q[q^{2} + \xi^{-2}]} \Bigr)
\nn && - \frac{1}{2\pi^{2}} \int_{0}^{\infty} {d \nu}
\coth\Bigl(\frac{\nu}{2T}\Bigr) \int_{0}^{\infty} d q q
\tan^{-1}\Bigl( \gamma_{a} \frac{\nu}{q^{3}} \Bigr) .  \eqa The
first term is due to the Fermi liquid contribution of holon
excitations, and the second is the correlation length term. The
third contribution results from $z = 1$ antiferromagnetic
fluctuations while the fourth term arises from $z = 3$ valance
bond excitations. The last term originates from $z = 3$ gauge
fluctuations.

Considering the specific heat coefficient given by $\gamma(T) =
\frac{C(T)}{T} = - \frac{\partial^{2} F_{LW}(T)}{\partial T^{2}}$,
we obtain its analytic expression in each regime, \bqa && \gamma(T
> \frac{(\gamma_{v}\xi^{3})^{-1}}{2}) \approx 2^{8/3} \Bigl(
\frac{1}{2\pi^{2}} + \frac{1}{8\pi}\Bigr) \Bigl( 2
\gamma_{v}^{2/3} + \gamma_{a}^{2/3} \Bigr) \nn && \Bigl[
\int_{0}^{\infty} {d y} \Bigl\{ - \frac{y^{5/3}}{\sinh^{2}y} +
\frac{y^{8/3} \coth y}{\sinh^{2} y} \Bigr\} \Bigr] T^{-1/3} , \nn
&& \gamma(T < \frac{(\gamma_{v}\xi^{3})^{-1}}{2}) \approx 2^{8/3}
\Bigl( \frac{1}{2\pi^{2}} + \frac{1}{8\pi}\Bigr) \gamma_{a}^{2/3}
\nn && \Bigl[ \int_{0}^{\infty} {d y} \Bigl\{ -
\frac{y^{5/3}}{\sinh^{2}y} + \frac{y^{8/3} \coth y}{\sinh^{2} y}
\Bigr\} \Bigr] T^{-1/3} , \eqa where only dominant contributions
are shown. As discussed previously, we have three regimes, (A) $T
< \frac{(\xi^{3}\gamma_{v})^{-1}}{2}$ where both antiferromagnetic
and valance bond fluctuations are gapped, (B)
$\frac{(\xi^{3}\gamma_{v})^{-1}}{2} < T < \frac{\xi^{-1}}{2}$
where only valance bond fluctuations are critical, and (C)
$\frac{\xi^{-1}}{2} < T$ where both antiferromagnetic and valance
bond fluctuations are critical. In the regime (A) contributions
from superspin fluctuations exhibit an exponential dependence of
temperature and ignored in the low energy limit. Dominant
contributions are driven by $z = 3$ critical gauge fluctuations,
resulting in $\gamma(T) \sim T^{-1/3}$. In the regime (B)
antiferromagnetic fluctuations cause an exponential dependence of
temperature while both valance bond and gauge fluctuations give
rise to $\gamma(T) \sim T^{-1/3}$ due to their $z = 3$
criticality. In the regime (C) $z = 3$ critical valance bond
excitations and gauge fluctuations allow $\gamma(T) \sim T^{-1/3}$
while $z = 1$ critical antiferromagnetic fluctuations result in
$\gamma_{AF}(T>\frac{\xi^{-1}}{2}) = \frac{6}{\pi} \Bigl(
\int_{0}^{\infty} d x \frac{x^{3}}{\sinh^{2} x}\Bigr) T$,
sub-leading thus ignored in the low energy limit.

\subsection{Transport}

We discuss electrical transport of the effective field theory Eq.
(10). As shown in the Eliashberg equations (17), the holon
self-energy is given by two contributions \bqa &&
\Sigma_{\eta}^{v}(i\omega) = m_{\eta}^{2} T \sum_{i\Omega} \int
\frac{d^{d}q}{(2\pi)^{d}} \nn && \sum_{m,n=4}^{5}
D_{v}^{mn}(q,i\Omega) \gamma_{m} G_{\eta}(k+q,i\omega+i\Omega)
\gamma_{n} , \nn && \Sigma_{\eta}^{a}(i\omega) = T \sum_{i\Omega}
\int \frac{d^{d}q}{(2\pi)^{d}} \nn && D_{a}^{\mu\nu}(q,i\Omega)
\gamma_{\mu}\tau_{3} G_{\eta}(k+q,i\omega+i\Omega)
\gamma_{\nu}\tau_{3} ,   \eqa where the first is due to valance
bond or monopole fluctuations, and the second comes from gauge
fluctuations.

We first consider the self-energy due to valance bond
fluctuations. The above expression can be written as follows \bqa
&& \Sigma_{\eta}^{v}(i\omega) \approx - \frac{2 g m_{\eta}^{2}
}{\mu_{h}} T \sum_{i\Omega} \int \frac{d^{d}q}{(2\pi)^{d}} \nn &&
\frac{1}{ q^{2} + \xi^{-2} + \gamma_{v} \frac{|\Omega|}{q} }
\frac{i \gamma_{0} \omega + i \gamma_{i} k_{i}^{F} + \mu_{h}
\gamma_{0}} { i \omega + i \Omega - q\cos\theta} , \eqa where the
self-energy correction of doped holes via $z = 3$ valance bond
fluctuations is clearly seen. Here, $k_{F} = \sqrt{k_{x}^{F2} +
k_{y}^{F2}} = \mu_{h}$ is the holon Fermi momentum. Then, the
imaginary part of the self-energy is given by \bqa && \Im
\Sigma_{\eta}^{v}(\omega+i\delta) = \frac{g m_{\eta}^{2}
}{2\pi^{3} \mu_{h}} [\gamma_{0} \omega + i \gamma_{i} k_{i}^{F} +
\mu_{h} \gamma_{0}] \nn && \int_{0}^{|\omega|} {d\Omega_{1}}
\int_{\xi^{-1}}^{\infty} d q \frac{q}{\sqrt{q^{2} - (\omega +
\Omega_{1})^{2}}}\frac{\gamma_{v} \Omega_{1}q}{ q^{6} +
\gamma_{v}^{2}\Omega_{1}^{2} } , \nonumber \eqa where Wick
rotation is performed at zero temperature in order to see
frequency dependence of the self-energy.

Performing momentum and frequency integrals, we find \bqa && \Im
\Sigma_{\eta}^{v}(\omega > \frac{(\gamma_{v}\xi^{3})^{-1}}{2})
\approx \frac{g m_{\eta}^{2} }{4\sqrt{3}\pi^{2} \gamma_{v}^{1/3}}
\gamma_{0} |\omega|^{2/3} , \nn && \Im \Sigma_{\eta}^{v}(\omega <
\frac{(\gamma_{v}\xi^{3})^{-1}}{2}) \approx \frac{g m_{\eta}^{2}
\xi}{\sqrt{3} \pi^{2} } \gamma_{0} \omega + \gamma_{0}
\mathcal{O}(\omega^{2}) . \eqa Note that the $|\omega|^{2/3}$
behavior is the hallmark of $z = 3$ criticality in two
dimensions.\cite{Gauge_NFL} The self-energy correction due to
gauge fluctuations also gives rise to $\Im
\Sigma_{\eta}^{a}(\omega) \propto |\omega|^{2/3}$.

At finite temperatures the zero-frequency self-energy corrections
turn out to diverge in the one-loop approximation. However, such
divergences due to both gauge and valance bond fluctuations need
not be given much attention because such self-energies are not
gauge-invariant, thus they do not have any physical meaning. These
divergences should be considered as an artifact of gauge
non-invariance. Gauge invariance can be incorporated via vertex
corrections, which cancel the divergent parts in the
self-energies, giving rise to gauge invariant finite
contributions.\cite{Nambu,YBKim,DonHKim} This corresponds to the
transport time, given by $q^{2} \sim T^{\frac{2}{z}}$
multiplication in the quasiparticle life time. As a result, we
find the following expression for the electrical resistivity \bqa
&& \rho(T) \propto T^{4/3} , \eqa consistent with the previous
results.\cite{Gauge_NFL} We obtain non-Fermi liquid physics in
both quantum critical and disordered phases, where both valance
bond and gauge fluctuations cause the non-Fermi liquid transport
in the quantum critical regime while only gauge fluctuations
result in that in the disordered phase.

\section{Application to one dimension}

Although our effective field theory Eq. (10) is physically
reasonable, we would like to justify its validity applying it to
one dimension. In one dimension the spin sector is described by
the SO(4) WZW theory, and the charge sector is represented by
QED$_{2}$ without the chemical potential term. Accordingly, the
coupling term between valance bond fluctuations and holons is
adjusted. The resulting effective field theory in one dimension is
obtained to be \bqa && S = \int {d^{2} x} \Bigl[ \frac{1}{2g}
\sum_{k=1}^{4}(\partial_{\mu} v_{k})^{2} \nn && + i
\frac{2\pi}{Area(S^{3})} \int_{0}^{1}{dt} \epsilon_{abcd} v_{a}
\partial_{t} v_{b} \partial_{\tau} v_{c}
\partial_{x} v_{d} \Bigr] \nn && + \int {d^{2}
x} \Bigl[ \bar{\eta}\gamma_{\mu}(\partial_{\mu} -
ia_{\mu}^{3}\tau_{3} - iA_{\mu})\eta - m_{\eta} \bar{\eta} (i
\gamma_{5} v_{4} ) \eta \nn && +
\frac{1}{2e^{2}}(\epsilon_{3\mu\nu}\partial_{\mu}a_{\nu}^{3})^{2}
\Bigr]  , \eqa where $\gamma_{\mu}$ and $\gamma_{5}$ are $2\times
2$ matrices, and $\eta$ is a four component Dirac spinor. The
SO(4) WZW theory of the spin sector has been derived in Ref.
\cite{Tanaka_SO4}, using the path integral formulation for
non-abelian chiral anomaly. Here, we investigate the role of
massless Dirac fermions, introduced from Eq. (10) by reducing the
two dimensional theory down to one dimension.

Performing the abelian bosonization for the fermion sector, we
obtain the following expression \bqa && S = \int {d^{2} x} \Bigl[
\frac{1}{2g} \sum_{k=1}^{4}(\partial_{\mu} v_{k})^{2} \nn && + i
\frac{2\pi}{Area(S^{3})} \int_{0}^{1}{dt} \epsilon_{abcd} v_{a}
\partial_{t} v_{b} \partial_{\tau} v_{c}
\partial_{x} v_{d} \Bigr] \nn && + \int {d^{2} x} \Bigl[
\frac{1}{2}(\partial_{\mu}\phi_{+})^{2} +
\frac{1}{2}(\partial_{\mu}\phi_{-})^{2} \nn && + \Bigl(
\frac{\Lambda}{\pi} m_{\eta} \Bigr) v_{4} \sin \Bigl(
\sqrt{4\pi}\phi_{+} \Bigr) + \Bigl( \frac{\Lambda}{\pi} m_{\eta}
\Bigr) v_{4} \sin \Bigl( \sqrt{4\pi}\phi_{-} \Bigr) \nn && -
ia_{\mu}^{3}
\Bigl(\frac{1}{2\pi}\epsilon_{\mu\nu}\partial_{\nu}\phi_{+} -
\frac{1}{2\pi}\epsilon_{\mu\nu}\partial_{\nu}\phi_{-}\Bigr) \nn &&
-
iA_{\mu}\Bigl(\frac{1}{2\pi}\epsilon_{\mu\nu}\partial_{\nu}\phi_{+}
+ \frac{1}{2\pi}\epsilon_{\mu\nu}\partial_{\nu}\phi_{-}\Bigr) +
\frac{1}{2e^{2}}(\epsilon_{\mu\nu}\partial_{\mu}a_{\nu}^{3})^{2}
\Bigr] , \nn \eqa where the subscript $\pm$ in the bosonic field
$\phi_{\pm}$ represents the SU(2) doublet of $\tau_{3}$, and
$\Lambda$ is a cutoff associated with band linearization.

Performing integration for U(1) gauge fields, we find a mass-type
term $\frac{e^{2}}{8\pi^{2}}(\phi_{+} - \phi_{-})^{2}$. This
allows us to set $\phi_{+} = \phi_{-} \equiv \phi$ in the low
energy limit. Shifting $\sqrt{4\pi} \phi$ with $- \frac{\pi}{2} +
\sqrt{4\pi} \theta$, we are led to \bqa && S = \int {d^{2} x}
\Bigl[ \frac{1}{2g} \sum_{k=1}^{4}(\partial_{\mu} v_{k})^{2} \nn
&& + i \frac{2\pi}{Area(S^{3})} \int_{0}^{1}{dt} \epsilon_{abcd}
v_{a}
\partial_{t} v_{b} \partial_{\tau} v_{c}
\partial_{x} v_{d} \Bigr] \nn && + \int {d^{2} x} \Bigl[ (\partial_{\mu}\theta)^{2}
- \Bigl( \frac{2\Lambda}{\pi} m_{\eta} \Bigr) v_{4} \cos \Bigl(
\sqrt{4\pi}\theta \Bigr) \nn && -
iA_{\mu}\Bigl(\frac{1}{\pi}\epsilon_{\mu\nu}\partial_{\nu}\theta
\Bigr) \Bigr] . \eqa It is interesting to see that valance bond
excitations lead to charge density wave fluctuations, consistent
with our expectation.

The valance bond and charge density wave coupling term can be
taken into account in the cumulant expansion, and the correction
part is given by \bqa && \delta S = - \frac{1}{2} \Bigl( \langle
S_{int}^{2} \rangle - \langle S_{int} \rangle^{2} \Bigr) \nn && =
- \frac{1}{2} \Bigl( \frac{2\Lambda}{\pi} m_{\eta} \Bigr)^{2} \int
d^{2}x \int d^{2}x' \nn && \Bigl\{ v_{4}(x) \langle \cos \Bigl(
\sqrt{4\pi}\theta(x) \Bigr) \cos \Bigl( \sqrt{4\pi}\theta(x')
\Bigr) \rangle v_{4}(x') \nn && + \cos \Bigl( \sqrt{4\pi}\theta(x)
\Bigr) \langle v_{4}(x)v_{4}(x') \rangle \cos \Bigl(
\sqrt{4\pi}\theta(x') \Bigr) \Bigr\} \nn && \equiv \delta
S_{v_{4}} + \delta S_{\theta} , \eqa where $S_{int} = - \int d^{2}
x \Bigl( \frac{2\Lambda}{\pi} m_{\eta} \Bigr) v_{4} \cos \Bigl(
\sqrt{4\pi}\theta \Bigr)$ is the coupling term.

It is not difficult to evaluate the density-density correlation
function since charge fluctuations are described by the
noninteracting Gaussian ensemble if metallic charge dynamics is
assumed. In this case we find \bqa && \langle \cos \Bigl(
\sqrt{4\pi}\theta(x) \Bigr) \cos \Bigl( \sqrt{4\pi}\theta(x')
\Bigr) \rangle \propto \cosh \Bigl( 4\pi \langle \theta(x)
\theta(x') \rangle \Bigr) \nn && = \cosh \Bigl( 4\pi
\mathcal{C}_{\theta} \ln |x - x'| \Bigr) \rightarrow |x -
x'|^{4\pi\mathcal{C}_{\theta}} , \eqa where $\mathcal{C}_{\theta}$
is a positive numerical constant, and the last part is valid at
large distances, i.e., $|x-x'|\rightarrow\infty$.

Inserting this expression into the spin sector, we obtain an
effective theory for SO(4) spin fluctuations \bqa && S_{v_{4}} =
\int {d^{2} x} \Bigl[ \frac{1}{2g} \sum_{k=1}^{4}(\partial_{\mu}
v_{k})^{2} \nn && + i \frac{2\pi}{Area(S^{3})} \int_{0}^{1}{dt}
\epsilon_{abcd} v_{a}
\partial_{t} v_{b} \partial_{\tau} v_{c}
\partial_{x} v_{d} \Bigr] \nn && - \int d^{2}x \int d^{2}x'
\mathcal{C}_{v_{4}} \Bigl( \frac{2\Lambda}{\pi} m_{\eta}
\Bigr)^{2} v_{4}(x) |x - x'|^{4\pi\mathcal{C}_{\theta}} v_{4}(x')
, \nn \eqa where $\mathcal{C}_{v_{4}}$ is a positive numerical
constant. An important point is that metallic charge fluctuations
give rise to confining interactions between monopoles, suppressing
monopole fluctuations. This tendency is completely consistent with
our previous two dimensional analysis, where holon fluctuations
cause dissipative monopole dynamics described by $z = 3$,
prohibiting their proliferation. In one dimension charge dynamics
suppresses monopole condensation more strongly.

An immediate question is the nature of spin dynamics descried by
Eq. (41). Since monopole excitations will be suppressed via charge
fluctuations, spinon deconfinement is expected to appear. A
relevant point is whether spin fluctuations are critical or not.
The SO(4) WZW theory is well known to exhibit criticality without
the fermion-induced monopole-suppressing term at half filling,
where monopole excitations turn out to be irrelevant due to the
presence of the topological WZW term, which differs completely
from the charge-dynamics monopole-suppressing mechanism. Since the
SO(4) symmetry is broken by the presence of charge fluctuations
shown by the last term of Eq. (41), we expect that the WZW term
may not be relevant in this case, thus resulting in spin gap. This
seems to be consistent with our physical intuition that charge
fluctuations will cut spin correlations, making their correlation
length short. Then, the resulting state is identified with the
Luther-Emery phase, where spin fluctuations are gapped and charge
excitations exhibit enhanced superconducting
correlations.\cite{Luther_Emery}

On the other hand, if charge fluctuations are gapped, i.e., in the
Mott insulating phase, their density-density correlations will
vanish at large distances as follows, $\langle \cos \Bigl(
\sqrt{4\pi}\theta(x) \Bigr) \cos \Bigl( \sqrt{4\pi}\theta(x')
\Bigr) \rangle \propto e^{- |x-x'|/\xi_{\eta}}$, where
$\xi_{\eta}^{-1}$ is associated with their excitation gap. Then,
spin dynamics will be described by the pure SO(4) WZW theory in
the long wave-length limit. As a result, the critical spin liquid
Mott insulator is expected to appear in this case.

\section{Superconductivity and stability of the anomalous metal}

Until now, we have seen the nature of the anomalous metallic
phase. An important problem is how d-wave superconductivity
evolves from the non-Fermi liquid state. In our effective theory
approach superconductivity can appear only from pairing of
fermionic doped holes. As shown explicitly in our effective action
Eq. (10), interactions between doped holes are mediated by both
gauge and valance bond fluctuations. An important thing is that
these interactions compete with each other.

Since holons carry an isospin quantum number represented by the
$\tau_{3}$ matrix, two kinds of fermion pairs can be considered.
When their isospin quantum numbers are different from each other,
gauge fluctuations cause attractive interactions for the
particle-particle channel while valance bond fluctuations give
rise to repulsive ones. This is nothing but the pairing
possibility between $b_{1}$ and $b_{2}$ bosons in the bosonic
description\cite{SU2_EFT} of the SU(2) slave-boson theory,
discussed by Wen and Lee.\cite{Lee_Nagaosa_Wen} In this respect
such pairing possibility is the unique feature of the SU(2)
slave-boson description for superconductivity.

On the other hand, when their isospin quantum numbers are the same
as each other, gauge fluctuations give rise to repulsive
interactions for the particle-particle channel while valance bond
fluctuations cause attractive ones between holons on different
sublattices. Do not confuse the present uniform (ferromagnetic)
gauge fluctuations with staggered (antiferromagnetic) ones.
Staggered gauge fluctuations, discussed previously in the
slave-fermion context, cause attractive interactions between doped
holes on different sublattices. Gauge fluctuations appearing in
the present theory are in the uniform channel associated with the
third component of the SU(2) slave-boson theory. The presence of
$\tau_{3}$ matrix in the gauge coupling supports this fact. As a
result, such holons have the same gauge charges, repulsing each
other.

Based on the above discussion, we conclude that emergence of
superconductivity is determined by the relative strength between
the couplings of gauge and valance bond fluctuations with doped
holes. An interesting observation is that the presence of
repulsive interactions in both cases may favor d-wave pairing of
doped holes.

Now, we derive a full set of Eliashberg equations including the
pairing channel. In this paper we consider only the first case,
that is, pairing between different isospins. One can construct the
Luttinger-Ward functional in the same way as the previous case Eq.
(16), \bqa && F_{LW} = F_{LW}^{\eta} + F_{LW}^{v} + F_{LW}^{a} +
Y_{v} + Y_{a} , \nn && F_{LW}^{\eta} = - T \sum_{i\omega}
\int\frac{d^{d}k}{(2\pi)^{d}} \mathbf{tr} \Bigl[\ln\Bigl\{
g_{\eta}^{-1}(k,i\omega) \mathbf{I} +
\mathbf{\Sigma_{\eta}}(i\omega) \Bigr\} \nn && -
\mathbf{\Sigma_{\eta}}(i\omega) \mathbf{G_{\eta}}(k,i\omega)
\Bigr] , \nn && F_{LW}^{v} = T \sum_{i\Omega}
\int\frac{d^{d}q}{(2\pi)^{d}} \Bigl[ \sum_{m,n=1}^{5} \ln\Bigl\{
d_{v}^{-1}(q,i\Omega)\delta_{mn} \nn && + \Pi_{v}^{mn}(q,i\Omega)
\Bigr\} - \sum_{m,n=1}^{5}
\Pi_{v}^{mn}(q,i\Omega)D_{v}^{mn}(q,i\Omega)\Bigr] - m_{v}^{2} ,
\nn && F_{LW}^{a} = T \sum_{i\Omega} \int\frac{d^{d}q}{(2\pi)^{d}}
\Bigl[\ln\Bigl\{ d_{a}^{-1}(q,i\Omega) + \Pi_{a}(q,i\Omega)
\Bigr\} \nn && - \Pi_{a}(q,i\Omega) D_{a}(q,i\Omega) \Bigr] , \nn
&& Y_{v} = - \frac{m_{\eta}^{2}}{2} T \sum_{i\omega}
\int\frac{d^{d}k}{(2\pi)^{d}} T \sum_{i\Omega}
\int\frac{d^{d}q}{(2\pi)^{d}} \nn && \sum_{m,n=4}^{5} \mathbf{tr}[
D_{v}^{mn}(q,i\Omega) \gamma_{m}
\mathbf{G_{\eta}}(k+q,i\omega+i\Omega)
\gamma_{n}\mathbf{G_{\eta}}(k,i\omega)] , \nn && Y_{a} = -
\frac{1}{2} T \sum_{i\omega} \int\frac{d^{d}k}{(2\pi)^{d}} T
\sum_{i\Omega} \int\frac{d^{d}q}{(2\pi)^{d}} \nn && \mathbf{tr} [
D_{a}^{\mu\nu}(q,i\Omega) \gamma_{\mu}\tau_{3}
\mathbf{G_{\eta}}(k+q,i\omega+i\Omega)
\gamma_{\nu}\tau_{3}\mathbf{G_{\eta}}(k,i\omega)] . \eqa Here,
$\mathbf{G_{\eta}}^{-1}(k,i\omega) = \mathbf{g}^{-1}(k,i\omega) +
\mathbf{\Sigma}(k,i\omega)$ is the matrix green's function with an
anomalous propagator, where $\mathbf{g} (k,i\omega) = g(k,i\omega)
\mathbf{I}$ with ${g}^{-1}(k,i\omega) = (i \gamma_{\mu} k_{\mu} +
\mu_{h}\gamma_{0})^{-1}$ is the bare green's function and
$\mathbf{\Sigma}(k,i\omega)
= \left( \begin{array}{cc} \Sigma(k,i\omega) & \Delta(k,i\omega) \\
\Delta^{*}(k,i\omega) & \Sigma(k,i\omega) \end{array}\right)$ is
the self-energy matrix with its anomalous part.

The above Luttinger-Ward functional results in the following
Eliashberg equations \bqa && \mathbf{\Sigma_{\eta}}(i\omega) =
m_{\eta}^{2} T \sum_{i\Omega} \int \frac{d^{d}q}{(2\pi)^{d}}
\sum_{m,n=4}^{5} \nn && D_{v}^{mn}(q,i\Omega) \gamma_{m}
\mathbf{G_{\eta}}(k_{F}+q,i\omega+i\Omega) \gamma_{n} + T
\sum_{i\Omega} \int \frac{d^{d}q}{(2\pi)^{d}} \nn &&
D_{a}^{\mu\nu}(q,i\Omega) \gamma_{\mu}\tau_{3}
\mathbf{G_{\eta}}(k_{F}+q,i\omega+i\Omega) \gamma_{\nu}\tau_{3} ,
\nn && \Pi_{v}^{mn}(q,i\Omega) = T \sum_{i\omega}\int
\frac{d^{d}k}{(2\pi)^{d}} \nn && \sum_{m,n=4}^{5} \Bigl( -
\frac{m_{\eta}^{2}}{2} \mathbf{tr}
[\gamma_{m}\mathbf{G_{\eta}}(k+q,i\omega+i\Omega)\gamma_{n}\mathbf{G_{\eta}}(k,i\omega)]
\Bigr) , \nn && \Pi_{a}^{\mu\nu}(q,i\Omega) = T \sum_{i\omega}
\int\frac{d^{d}k}{(2\pi)^{d}} \nn && \Bigl( - \frac{1}{2}
\mathbf{tr} [ \gamma_{\mu}\tau_{3}
\mathbf{G_{\eta}}(k+q,i\omega+i\Omega)
\gamma_{\nu}\tau_{3}\mathbf{G_{\eta}}(k,i\omega)] \Bigr) , \eqa
where this is basically the same as the previous one Eq. (17), but
the holon propagator is replaced with a matrix including its
superconducting part. If we rewrite the above equations with each
component of the holon propagator, we obtain \bqa &&
\Sigma(k,i\omega) \nn && = m_{\eta}^{2} T\sum_{i\Omega,q} \sum_{n
= 3}^{4} \gamma_{n} G(k+q,i\omega+i\Omega) D_{nn}^{v}(q,i\Omega)
\gamma_{n} \nn && + T\sum_{i\Omega,q} \sum_{i,j=x,y}\gamma_{i}
G(k+q,i\omega+i\Omega) D_{ij}^{a}(q,i\Omega) \gamma_{j} \eqa and
\bqa && \Delta(k,i\omega) \nn && = m_{\eta}^{2} T\sum_{i\Omega,q}
\sum_{n = 3}^{4} \gamma_{n} F(k+q,i\omega+i\Omega)
D_{nn}^{v}(q,i\Omega) \gamma_{n} \nn && - T\sum_{i\Omega,q}
\sum_{i,j=x,y}\gamma_{i} F(k+q,i\omega+i\Omega)
D_{ij}^{a}(q,i\Omega) \gamma_{j}  , \eqa where the normal and
abnormal holon propagators are given by \bqa && G(k,i\omega) =
\frac{{g}^{-1}(k,i\omega) +
\Sigma(k,i\omega)}{[{g}^{-1}(k,i\omega) + \Sigma(k,i\omega)]^{2} -
|\Delta(k,i\omega)|^{2}} , \nn && F(k,i\omega) = -
\frac{\Delta(k,i\omega)}{[{g}^{-1}(k,i\omega) +
\Sigma(k,i\omega)]^{2} - |\Delta(k,i\omega)|^{2}} . ~~~~~~~~ \eqa
As shown in Eqs. (45) and (46), gauge fluctuations cause
attractive interactions between holons with different isospins,
guaranteed by the the presence of the $\tau_{3}$ matrix, while
valance bond fluctuations result in repulsive ones. The presence
of repulsive interactions opens the possibility of d-wave pairing
of doped holes.

As discussed above, there is another pairing channel between
holons with same isospins but on different sublattices, induced by
valance bond fluctuations. Comparison of these two superconducting
phases by solving each Eliashberg equations remains as an
important future work. What we would like to mention is that
because both gauge and valance bond fluctuations give rise to
competing interactions, the present anomalous metallic phase is
expected to be stable against superconductivity in some parameter
regions of the effective theory Eq. (10). Even if such a metallic
phase becomes unstable against superconductivity at zero
temperature, at finite temperatures the present non-Fermi liquid
physics would survive.

There is another instability channel associated with particle-hole
pairing. Different from the particle-particle channel, gauge
fluctuations cause attractive interactions for the particle-hole
channel when holons have the same isospin. As pointed out by
Altshuler et al.,\cite{2kF_instability} the $2k_{F}$ vertex with
the Fermi momentum $k_{F}$ is power-law diverging with its
exponent $1/N$ approximately, where $N$ is the flavor number of
fermions. They showed that divergence of particle-hole
susceptibility depends on the exponent of the diverging vertex. In
other words, the particle-hole pairing instability arises when the
divergence of the $2k_{F}$ interaction vertex is sufficiently
strong. In this respect the present effective theory with the
large $N$ approximation is expected to be free from such
particle-hole instabilities. Remember that the large $N$
approximation is consistent with the self-consistent Eliashberg
framework. Interestingly, valance bond fluctuations are against
such particle-hole pairings as they compete with gauge
fluctuations for superconductivity. More quantitative analysis for
the particle-hole channel is required when the flavor number of
fermions is not large. However, this is beyond the scope of the
present paper since it is not clear whether even the Eliashberg
approximation is stable or not in this case.

In the recent publication\cite{Two_Leg_Ladder} the presence of an
interesting inhomogeneous superconducting state was demonstrated
in the two-leg ladder system based on the renormalized mean-field
theory, where a heavy numerical analysis was performed for solving
self-consistent mean-field equations of order parameters in real
space. From the present approach it is difficult to see the
emergence of such complicated inhomogeneous order parameter
patterns because our effective field theory is based on the
uniform phase of order parameters.

Applying the present theoretical framework to the two-leg ladder
system, one would see that gauge fluctuations enhance the $2k_{F}$
particle-hole vertex to cause power-law divergence, but the
particle-hole susceptibility does not diverge at least in the
large $N$ approximation, as discussed above. However, charge and
spin density waves can certainly occur when the fermion flavor
number is small. In this respect, when the coupling strength
between doped holes and valance bond fluctuations is large but the
fermion flavor number is not large, an inhomogeneous
superconducting phase is expected to appear. Unfortunately, fully
self-consistent analysis including both particle-particle and
particle-hole instabilities is clearly beyond the scope of the
present paper.

\section{Summary}

Fermionizing the charge sector and bosonizing the spin part in the
SU(2) slave-boson theory, we have derived an effective field
theory for dynamics of doped holes in the antiferromagnetically
correlated spin background, where spin fluctuations are described
by the SO(5) WZW theory while charge dynamics is expressed by
non-relativistic QED$_{3}$ around four Dirac nodes. In particular,
hole dynamics affects deconfinement of bosonic spinons in the
SO(5) WZW theory through the coupling term between valance bond
(monopole) fluctuations and fermionic holons. Such interactions
give rise to $z = 3$ criticality for monopole dynamics,
prohibiting their proliferation in the presence of the WZW term.
As a result, holon fluctuations turn out to help spin
fractionalization near the quantum critical point.

We have investigated thermodynamics and transport in the
Eliashberg framework for our effective field theory. We pointed
out that the Eliashberg framework is the lowest order
self-consistent approximation well controlled in our effective
field theory, where Migdal theorem works well owing to $z = 3$
criticality, and the large $N_{\eta}$ limit with the number of
Dirac nodes $N_{\eta}$ is naturally allowed. We find that spin
fluctuations are described by $z = 1$ for antiferromagnetic
fluctuations and $z = 3$ for valance bond excitations, giving rise
to three regimes, where superspin fluctuations are gapped at low
temperatures, only valance bond excitations are critical at
intermediate temperatures, and superspin fluctuations are critical
at high temperatures. Both valance bond and gauge fluctuations are
described by $z = 3$ critical theory, and we find non-Fermi liquid
physics for thermodynamics and electrical transport near the
quantum critical point, consistent with $z = 3$ scaling. In
addition, even in the quantum disordered phase such non-Fermi
liquid physics is preserved owing to critical gauge fluctuations.

To further justify our effective field theory, we have applied it
to one dimension, physically well known. In one dimension spin
fluctuations are described by the SO(4) WZW theory while charge
excitations are represented by QED$_{2}$. We have taken the
monopole-holon coupling term into account in the abelian
bosonization framework. We demonstrated that holon dynamics
results in confining interactions between monopole excitations.
Thus, we conclude that charge fluctuations help spinon
deconfinement in both one and two dimensional cases while charge
dynamics suppresses monopole fluctuations more strongly in one
dimension.

We have discussed stability of the non-Fermi liquid metallic phase
against superconductivity and density waves. An interesting
observation is that two kinds of holon pairing channels exist due
to the presence of the isospin quantum number in the SU(2)
slave-boson description. In the different isospin channel
attractive pairing interactions are caused by gauge fluctuations
while in the same isospin channel such interactions arise from
valance bond fluctuations. An important thing is that such gauge
and valance bond interactions compete with each other. As a
result, superconductivity is expected to appear in a limited
parameter range. The presence of repulsive interactions allows the
possibility of d-wave pairing of doped holes. For the
particle-hole channel, we argued that as far as the fermion flavor
number is sufficiently large, consistent with the Eliashberg
approximation, the homogeneous metallic phase can be stable
against charge and spin density waves.

We would like to emphasize that the SU(2) structure is important
for our treatment. If we start from the U(1) slave-boson
representation, the fermionization procedure is not performed
naturally since we have nonzero net flux owing to the presence of
finite density of holons. Furthermore, the SO(5) WZW theory for
the spin sector does not appear owing to the contribution from a
chemical potential term in the U(1) slave-boson framework.

Our effective field theory exhibits direct interactions between
monopoles and holes, associated with deconfinement of bosonic
spinons away from half filling. The important issue how doped
holes affect spinon deconfinement deserves to be studied more
carefully.

We appreciate extremely helpful discussions with A. Tanaka for his
SO(5) WZW description. This work is supported by the French
National Grant ANR36ECCEZZZ. K.-S. Kim is also supported by the
Korea Research Foundation Grant (KRF-2007-357-C00021) funded by
the Korean Government.


\begin{thebibliography}{9}
\bibitem{U1_Superfluid} The U(1) slave-boson formulation cannot
explain the doping independent decreasing ratio of superfluid
weight owing to charge renormalization of nodal quasiparticles as
a function of hole concentration.\cite{Lee_Nagaosa_Wen} In
addition, it cannot explain the existence of coherent electron
excitations near the nodal points in the anomalous normal state
although instanton physics of U(1) gauge fields is argued to cause
confinement and result in such excitations.\cite{Lee_Nagaosa_Wen}
\bibitem{SO5} In the pure SO(5) nonlinear $\sigma$ model we find
\bqa && 1 = \frac{5g}{\pi} T \Bigl\{ \ln \sinh\Bigl(
\frac{\Lambda}{2T} \Bigr) - \ln \sinh\Bigl(
\frac{\xi^{-1}_{v}}{2T} \Bigr) \Bigr\} \nonumber \eqa for the
correlation length. Its solution is given by \bqa &&
\xi_{v}^{-1}(g \sim g_{c}^{v}; T < \frac{\xi^{-1}_{v}}{2}) \approx
\Lambda - \frac{2\pi}{5g} \equiv \frac{2\pi}{5}
\Bigl(\frac{1}{g_{c}^{v}} - \frac{1}{g} \Bigr) , \nn &&
\xi_{v}^{-1}(g \sim g_{c}^{v}; T > \frac{\xi^{-1}_{v}}{2}) = 2 T
e^{\frac{\pi}{5}\frac{ \Bigl(\frac{1}{g_{c}^{v}} - \frac{1}{g}
\Bigr)}{T} } , \nonumber \eqa resulting in $g_{c}^{v} > g_{c}$,
consistent with our intuition.
\end{thebibliography}
\end{document}